\documentclass{llncs}

\usepackage{epsfig}
\usepackage{amsfonts}

\newcommand{\F}{\mbox{${\cal F}$}}
\newcommand{\R}{\mbox{${\cal R}$}}
\newcommand{\A}{\mbox{${\cal A}$}}
\newcommand{\la}{\leftarrow}
\newcommand{\ra}{\rightarrow}

\newcommand{\norm}[1]{\mbox{$\parallel{#1}\parallel$}}
\newcommand{\level}[1]{\mbox{$|{#1}|$}}
\newcommand{\llevel}[1]{\mbox{$\parallel{#1}\parallel$}}

\begin{document}

\pagestyle{plain}

\title{\bf Termination Proofs for Logic Programs with Tabling\thanks{This 
article is a collection and integration of a number of results that 
appeared---sometimes in weaker forms---in the conference papers
\cite{Verbaeten_ppdp99} and \cite{Verbaeten_flops99}.}
}

\author{Sofie Verbaeten\thanks{Research Assistant of the Fund for
                Scientific Research---Flanders (Belgium)(F.W.O.).}\inst{1} 
        \and Danny De Schreye\thanks{Senior Research Associate of
                F.W.O.~Flanders.}\inst{1}
        \and Konstantinos Sagonas\inst{2} 
       }

\institute{Department of Computer Science \\
           K.~U.~Leuven, Celestijnenlaan 200A \\ 
           B-3001 Heverlee, Belgium \\
           \email{$\{$sofie,dannyd$\}$@cs.kuleuven.ac.be}
\and
           Computing Science Department \\
           Uppsala University \\
           Box 311, S-751 05 Uppsala, Sweden \\
           \email{kostis@csd.uu.se}
}

\maketitle

\begin{abstract}
Tabled logic programming is receiving increasing attention
in the Logic Programming community.
It avoids many of the shortcomings of SLD execution and provides
a more flexible and often extremely efficient execution mechanism for 
logic programs.
In particular, tabled execution of logic programs terminates more often
than execution based on SLD-resolution.
In this article, 
we introduce two notions of universal termination of
logic programming
with Tabling:
quasi-termination and (the stronger notion of) LG-termination.
We present sufficient conditions for these two notions of termination,
namely quasi-acceptability and LG-acceptability,
and we show that these conditions are also necessary in case the 
tabling is well-chosen.
Starting from these conditions, we give modular termination proofs,
i.e., proofs capable of combining termination proofs of separate programs
to obtain termination proofs of combined programs.
Finally, in the presence of mode information, 
we state sufficient conditions which form the basis for automatically
proving termination in a constraint-based way.
\end{abstract}

\section{Introduction}

Tabled logic programming 
\cite{ChenWarren:JACM96,BolDeg@ICLP-93,tamaki:sato,vieille} is 
receiving increasing attention
in the Logic Programming community.
It avoids many of the shortcomings of SLD(NF) execution and provides
a more flexible and often extremely efficient execution mechanism for 
logic programs.
Furthermore, tabled execution of logic programs terminates more often
than execution based on SLD-resolution.  In particular, all programs
that terminate under SLD also terminate under tabled execution.  So,
if a program can be proven to terminate under SLD-resolution (by one
of the existing automated techniques surveyed in~\cite{survey}), then
the program will trivially also terminate under SLG-resolution, the
resolution principle of tabling; see~\cite{ChenWarren:JACM96}.  But,
since there are SLG-terminating programs which are not
SLD-terminating, more effective proof techniques need to and can be
found.

The idea underlying \emph{tabling} is quite simple. Essentially, under
a tabled execution mechanism, answers for selected tabled atoms as
well as these atoms are stored in a table.  When an identical
(up to renaming of variables) 
such atom is recursively called, the selected atom is not resolved
against program clauses; instead, all corresponding answers computed
so far are looked up in the table and the corresponding answer
substitutions are applied to the atom. This process is repeated for
all subsequent computed answer substitutions that correspond to the
atom.

We study \emph{universal} termination of \emph{definite} tabled logic
programs executed under SLG-resolution using a fixed
\emph{left-to-right selection rule} (we drop the ``S'' in SLD and SLG
whenever we refer to the left-to-right selection rule).  We introduce
a first basic notion of termination under tabled execution, called
{\em quasi-termination}.  Quasi-termination captures the property
that, under an LD-computation, a given atomic query leads to only
finitely many different
non-variant
calls to tabled predicates and
there is no infinite derivation consisting of queries with only
selected non-tabled atoms.  In a broader context, the notion of
quasi-termination and techniques for proving it are of independent
interest; they can be used to e.g.\/ ensure termination of off-line
specialisation of logic programs, whether tabled or not;
see~\cite{BTA@ESOP-98}.
However, the notion of quasi-termination only partially corresponds to
our intuitive notion of a ``terminating computation''.  This is
because an atom can have infinitely many computed answers (which does
not have to lead to infinitely many new calls).  Therefore, we also
introduce the stronger notion of {\em LG-termination}.  A program $P$
LG-terminates w.r.t.\ a given atomic query iff $P$ quasi-terminates
w.r.t.\ the query and the set of all computed answers for calls in the
LD-computation of the query is finite.

We present \emph{sufficient conditions} for these two notions of
termination under tabled execution: namely, quasi-acceptability for
quasi-termination and LG-acceptability for LG-termination.  We show
that these conditions are also \emph{necessary} in case the set of
tabled predicates is \emph{well-chosen}; see Section~\ref{sectchar}.
Our termination conditions are adapted from the acceptability notion
for LD-termination defined in~\cite{framework}, and not from the more
``standard'' definition of acceptability by Apt and Pedreschi
in~\cite{apt:pedreschi}.  The reason for this choice is that the
quasi-termination as well as the LG-termination property of a tabled
program and query is \emph{not}~closed under substitution.  The
acceptability notion in~\cite{apt:pedreschi} is expressed in terms of
ground instances of clauses and its associated notion of
LD-termination is expressed in terms of the set of all queries that
are bounded under the given level mapping. Such sets are closed under
substitution. Because quasi-termination and LG-termination lack
invariance under substitution, we use a stronger notion of
acceptability, capable of treating \emph{any} set of queries.

Besides a characterisation of the two notions of universal termination
under tabled execution, we also give \emph{modular termination
conditions}, i.e., conditions on two programs $P$ and $R$, where $P$
extends $R$, ensuring termination of the union $P \cup R$.  Such
modular proofs were already motivated in the literature in the context
of termination under SLD-resolution (see for
instance~\cite{apt:modular}).  Indeed, for programming in the large,
it is important to have modular termination proofs, i.e., proofs that
are capable of combining termination proofs of separate programs to
obtain termination proofs of combined programs.

Finally, we present easy to \emph{automate}, sufficient conditions for
quasi-termi\-na\-tion and LG-termination.  To this end, we use mode
information: we consider simply moded, well-moded programs and
queries.  We point out how these termination conditions could be
automated, by extending the recently developed, constraint-based,
automatic termination analysis for SLD-resolution
of~\cite{stef:constraints}.

All the above mentioned results are developed and presented for a
\emph{mixed tabled/non-tabled} execution mechanism.
This means that, in the execution, only a subset of the predicates
(specified by the programmer) will be tabled, while standard
LD-resolution
steps are applied to all others.
In Section~\ref{sectproltab}, we 
discuss
the benefits of having such a mixed execution mechanism.
This focus on mixed execution considerably strengthens our results.
In particular, 
our results
both introduce new termination conditions for
(fully) tabled logic programs, and at the same time generalize
existing termination conditions for LD-resolution.
Of course, this choice also makes the results more technically
involved.

The rest of the article is structured as follows.  In
Section~\ref{sectprel}, we define some preliminary concepts, in
particular the notion of finitely partitioning level mapping, which
plays a central role in our termination conditions.  Next, in
Section~\ref{sectproltab}, we recall the execution mechanism of
LG-resolution, the tabled-based resolution strategy used in this
article.
We first present examples from context-free grammar recognition
and parsing which motivate the need to freely mix 
untabled and tabled execution
and then we formally define the 
resolution principle of tabling, called SLG-resolution. 
Next, in Section~\ref{sectter}, two notions of termination of 
LG-resolution
are introduced: quasi-termination 
and the stronger notion of LG-termination.
We also define a transformation on programs which reduces the problem
of proving LG-termination to the problem of proving quasi-termination.
In Section~\ref{sectchar},
sufficient (and also necessary in case the tabling is {\em well-chosen})
conditions for the two notions of termination
are given:
the condition of quasi-acceptability for quasi-termination
(Subsection~\ref{ssectquasichar})
and the condition of LG-acceptability for LG-termination
(Subsection~\ref{ssectlgchar}).
Modular termination conditions, i.e., conditions that are capable
of combining termination proofs of separate programs to obtain 
termination proofs of combined programs,
are given in Section~\ref{sectmod}: in Subsection~\ref{ssectquasimod} for
quasi-termination, and in Subsection~\ref{ssectlgmod} for
LG-termination.
In Subsection~\ref{ssectconstrlm}, more detailed modular termination
conditions for quasi-termination are given, which also provide an
incremental construction of an appropriate level mapping.
Finally, in Section~\ref{sectauto}, we investigate conditions for
termination of LG-resolution which are easy to automate.  In
particular, our eventual goal is to extend the constraint-based
automatic approach towards LD-termination of~\cite{stef:constraints},
in order to prove termination of tabled logic programs in an automatic
way.  Our extension is restricted to the class of simply moded,
well-moded programs and queries, which we recall
from~\cite{apt:etalle}.  Only quasi-termination is considered in
Section~\ref{sectauto}; the results for LG-termination carry over in
the same way.  We end with some concluding remarks, a discussion on
related work and with some topics for future research.

\section{Preliminaries} \label{sectprel}

We assume familiarity with the basic concepts of logic programming;
see~\cite{lloyd,apt}.
Throughout the article, $P$ will denote a definite logic program.
By $Pred_P$, $Fun_P$ and $Const_P$ we denote the set of predicate,
function and constant symbols occurring in $P$.
We assume that these sets are finite.
By $Def_P$ we denote the set of predicates defined in $P$
(i.e., predicates occurring in the head of a  clause of $P$). 
By $Rec_P$,  resp.\ $NRec_P$, 
we denote the set of (directly or indirectly)
recursive, resp.\ non-recursive, 
predicates of the program $P$ (so $NRec_P = Pred_P \setminus Rec_P$).
If $A = p(t_1,\ldots,t_n)$, then we denote by $Rel(A)$ 
the predicate symbol $p$ of $A$; i.e., $Rel(A) = p$.
We call $A = p(t_1,\ldots,t_n)$ a $p$-atom.

The {\em extended Herbrand Universe}, $U_P^E$, and the 
{\em extended Herbrand Base},
$B_P^E$, associated with a program $P$, were introduced in~\cite{FLM89}.
They are defined as follows. Let $Term_P$ and $Atom_P$ denote the set of 
respectively all terms and atoms that can be constructed from the alphabet
underlying $P$.
The variant relation, denoted $\approx$, defines an equivalence.
$U_P^E$ and $B_P^E$ are respectively the quotient sets 
$Term_P/\approx$ and $Atom_P/\approx$.
For any term $t$ (or atom $A$), we denote its class in $U_P^E$ ($B_P^E$)
as $\tilde{t}$ ($\tilde{A}$).
However, when no confusion is possible, we omit the tildes.
For $\Pi \subseteq Pred_P$, we denote with 
$B_{\Pi}^E$ the subset of $B_P^E$ consisting of 
(equivalence classes of) atoms based on the predicate symbols of $\Pi$.
So $B_P^E$ can be seen as an abbreviation of $B_{Pred_P}^E$.

Let $P$ be a program and $p, q \in Pred_P$.
We say that $p$ {\em refers to} $q$ in $P$ iff there is a clause in $P$ with
$p$ in the head and $q$ occurring in the body.
We say that $p$ {\em depends on} $q$ in $P$, and write
$p \sqsupseteq q$, iff $(p,q)$ is in the reflexive, transitive closure
of the relation refers to.
Note that, by definition, each predicate depends on itself. 
We write $p \simeq q$ iff $p \sqsupseteq q$, $q \sqsupseteq p$
($p$ and $q$ are mutually recursive or $p=q$).
The {\em dependency graph} $G_P$ of a program $P$ is a 
graph where the nodes are labeled with the predicates of $Pred_P$.
There is a {\em directed arc} from $p$ to $q$ in $G_P$ iff
$p$ refers to $q$.
A program $P$ {\em extends}
a program $R$ iff no predicate defined in $P$ occurs in $R$. 

As mentioned and used in the introduction, in analogy
with~\cite{apt:pedreschi}, we will refer to SLD-derivations
(see~\cite{lloyd}) following the left-to-right selection rule as
LD-derivations.  Other concepts adopt this naming accordingly.

\begin{definition}[call set associated to $S$] \label{defcallset}
Let $P$ be a program and
$S \subseteq B_P^E$. By $Call(P,S)$ we denote the subset of
$B_P^E$ such that $B \in Call(P,S)$ whenever a representant 
of $B$ is a selected
atom in an LD-derivation for some $P \cup \{\la A\}$, with 
$\tilde{A} \in S$.
\end{definition}

Throughout the article we assume that in any derivation of a query
w.r.t.\ a program,
representants of equivalence classes are systematically provided 
with fresh variables, to avoid the necessity of renaming apart.
In the sequel, we abbreviate most general unifier with $mgu$ and
LD-computed answer substitution with $cas$.

The concepts defined in the following 
Definitions \ref{defdirdes}, \ref{defdirsubder}
and \ref{defcallgraph},
will be used in the proofs of some theorems and propositions
of this article. 

\begin{definition}[direct descendant] \label{defdirdes}
Let $P$ be a program and $\tilde{A}, \tilde{B} \in B_P^E$.
We call $\tilde{B}$ a {\em direct descendant} of $\tilde{A}$ iff
there exists a clause $H \la B_1,\ldots,B_n$ in $P$ such that
$mgu(A,H)=\theta$ exists and,
there is an $i \in [1,n]$ such that there is an LD-refutation for
$\la (B_1,\ldots,B_{i-1})\theta$ with $cas$
$\theta_{i-1}$ and $B\approx B_i\theta\theta_{i-1}$.
\end{definition}

\begin{definition} [directed subsequence of an LD-derivation]
\label{defdirsubder}
Let $P$ be a program and $\tilde{A} \in B_P^E$. 
Let $\la A=G_0,$ $G_1,\ldots$ be an LD-derivation of $\la A$ in $P$.
A subsequence $G_{i_0},G_{i_1},\ldots$, with
$G_{i_j}=\la A_{i_j},\A_{i_j}$, is called a {\em directed
subsequence} iff for all $j \geq 0$,  
$\tilde{A}_{i_{j+1}}$ is a direct descendant of $\tilde{A}_{i_j}$ in 
the LD-derivation.
\end{definition}

\begin{definition}[call graph associated to $S$]
\label{defcallgraph}
Let $P$ be a program and $S \subseteq B_P^E$.
The {\em call graph $Call$-$Gr(P,S)$ associated to $P$ and $S$}
 is a graph such that:
 \begin{itemize}
 \item its set of nodes is $Call(P,S)$,
 \item there exists a directed arc from $\tilde{A}$ to $\tilde{B}$ iff
 $\tilde{B}$ is a direct descendant of $\tilde{A}$.
\end{itemize}
\end{definition}

We recall the definitions of norm and level mapping, which are useful
in the context of termination analysis (see \cite{survey} for a survey
on termination analyses for (S)LD-resolution).

\begin{definition}[norm]
A {\em norm} is a function $\norm{.} : U_P^E \rightarrow \bbbn$.
\end{definition}

\begin{definition}[level mapping]
A {\em level mapping} is a function $\level{.} : B_P^E \rightarrow
\bbbn$.
\end{definition}

A level mapping or norm is said to be {\em trivial} if it is the
constant $0$-mapping.

Our termination conditions are based on the following
concept of a finitely partitioning
level mapping.

\begin{definition}
[finitely partitioning level mapping] \label{deffinpartlm}
Let $P$ be a program and $C \subseteq B_{P}^E$.
A level mapping $\level{.}$ is {\em finitely partitioning on $C$}
iff for all $n \in \bbbn: \sharp( \level{.}^{-1}(n) \cap C ) < \infty$,
where $\sharp$ is the cardinality function.
\end{definition}

So, a level mapping $\level{.}$ is finitely partitioning on $C
\subseteq B_{P}^E$ if it does not map an infinite set of atoms of $C$
to the same natural number.
That is, $\level{.}$ {\em partitions} $C$ into {\em finite} subsets.
In particular, we have that every level mapping is finitely 
partitioning on a finite set $C$.

\section{Tabling in Logic Programs} \label{sectproltab}

Our experience is that tabled execution is used \emph{selectively} in
practice.  Thus,
before formally
defining the resolution principle of tabling, called SLG-resolution,
we first present some examples which motivate the need to freely mix
LD-resolution and tabled execution.

\subsection{Mixing Tabled and LD Execution: Motivating Examples} 
\label{ssectmotexa}

It has long been noted in the literature \cite{Early@CACM-70,PTQ},
that tabled evaluation can be used for context-free grammar
recognition and parsing: tabling eliminates redundancy and handles
grammars that would otherwise infinitely loop under Prolog-style
execution (e.g.\/ left-recursive ones).  The following program, where
all predicates are tabled, provides such an example.
\[
\left\{
\begin{array}{lll}  
  expr(Si,So) & \la &  expr(Si,S1),  S1 = ['+'|S2], term(S2,So) \\
  expr(Si,So) & \la &  term(Si,So) \\
  term(Si,So) & \la & term(Si,S1), S1 = ['*'|S2], primary(S2,So) \\
  term(Si,So) & \la & primary(Si,So) \\
  primary(Si,So) & \la & Si = ['('|S1], expr(S1,S2), S2 = [')'|So] \\
  primary(Si,So) & \la & Si = [I|So], integer(I)
\end{array}
\right.
\]
This grammar, recognizing arithmetic expressions containing additions
and multiplications over the integers, is left recursive---left
recursion is used to give the arithmetic operators their proper
associativity---and would be non-terminating for Prolog-style
execution.  Under tabled execution, left recursion is handled
correctly.  In fact, one only needs to table predicates $expr/2$ and
$term/2$ to get the desired termination behaviour; we can and will
safely drop the tabling of $primary/2$ in the sequel.
However, this integration of 
non-tabled (LD)
and tabled execution is perhaps a
trivial one.

To see why a non-trivial mix of tabled with LD execution is desirable
in practice, suppose that we want to extend the above recognition
grammar to handle exponentiation.  The most natural way to do so is to
introduce a new nonterminal, named $factor$, for handling
exponentiation and make it right recursive, since the exponentiation
operator is right associative.  The resulting grammar is as below
where only the predicates $expr/2$ and $term/2$ are tabled.
\[
\left\{
\begin{array}{lll}
  expr(Si,So) & \la & expr(Si,S1), S1 = ['+'|S2], term(S2,So) \\
  expr(Si,So) & \la & term(Si,So) \\
  term(Si,So) & \la & term(Si,S1), S1 = ['*'|S2], factor(S2,So) \\
  term(Si,So) & \la & factor(Si,So) \\
  factor(Si,So) & \la & primary(Si,S1), S1 = ['\land'|S2], factor(S2,So) \\
  factor(Si,So) & \la & primary(Si,So) \\
  primary(Si,So) & \la & Si = ['('|S1], expr(S1,S2), S2 = [')'|So] \\
  primary(Si,So) & \la & Si = [I|So] , integer(I)
\end{array}
\right.
\]
Note that, at least as far as termination is concerned, there is no
need to table the new nonterminal $factor$.  Indeed, Prolog's evaluation
strategy handles right recursion in grammars finitely.  In fact,
Prolog-style evaluation of right recursion is more efficient than its
tabled-based evaluation: Prolog has linear complexity for a simple
right recursive grammar, but with tabling implemented as in XSB the
evaluation could be quadratic as calls need to be recorded in the
tables using explicit copying.  Thus, it is important to allow tabled
and non-tabled predicates to be freely intermixed, and be able to
choose the strategy that is most efficient for the situation at hand.

By using tabling in context-free grammars, one gets a recognition
algorithm that is a variant of Early's algorithm (also known as active
chart recognition algorithm) whose complexity is polynomial in the
size of the input expression/string~\cite{Early@CACM-70}.  However,
often one wants to construct the parse tree(s) for a given input
string.  The usual approach is to introduce an extra argument to the
nonterminals of the input grammar---representing the portion of the
parse tree that each rule generates---and naturally to also add the
necessary code that constructs the parse tree.  This approach is
straightforward, but as noticed by Warren in~\cite{Warren-book}, using
the same program for recognition as well as parsing may be extremely
unsatisfactory from a complexity standpoint: in context-free grammars,
recognition is polynomial while parsing is exponential, since there can
be exponentially many parse trees for a given input string.  The
obvious solution is to use two interleaved versions of the grammar as
in the following program, which recognizes and parses the language
$a^nb$.
\[
\begin{array}{ll}
R: &
\left\{
\begin{array}{lll}
  s(Si,So) & \la & a(Si,S), S = [b|So] \\
  a(Si,So) & \la & a(Si,S), a(S,So) \\
  a(Si,So) & \la & Si = [a|So]
\end{array}
\right.
\\ ~ \\
P: &
\left\{
\begin{array}{lll}
  s(Si,So,PT) & \la & a(Si,S), S = [b|So], PT = spt(PTa,b), a(Si,S,PTa) \\
  a(Si,So,PT) & \la & a(Si,S), a(S,So), PT = apt(PT1,PT2), a(Si,S,PT1), \\
              &     & a(S,So,PT2)\\
  a(Si,So,PT) & \la & Si = [a|So] , PT = a
\end{array}
\right.
\end{array}
\]
Note that only $a/2$, i.e., the recursive predicate of the
`recognition' part, $R$, of the program (consisting of predicates
$s/2$ and $a/2$), needs to be tabled.  This action allows recognition
to terminate and to have polynomial complexity.  Furthermore, the
recognizer can now be used as a filter for the parsing process in the
following way: only after knowing that a particular part of the input
belongs to the grammar and having computed the exact substring that
each nonterminal spans, do we invoke the parsing routine on the
nonterminal to construct its (possibly exponentially many) parse
trees.  Doing so, avoids e.g.\ cases where it may take exponential
time to fail on an input string that does not belong in the given
language: an example for the grammar under consideration is the input
string $a^n$.  On the other hand, tabling the `parsing' part of the
program (consisting of predicates $s/3$ and $a/3$) does not affect the
efficiency of the process complexity-wise and incurs a small
performance overhead due to the recording of calls and their answers
in the tables.  Finally, note that the construction is modular in the
sense that the `parsing' part of the program, $P$, depends on the
`recognition' part, $R$, but not vice versa; we say that $P$
\emph{extends} $R$.

\subsection{SLG-Resolution} \label{ssectslg}
In this article, we consider termination of SLG-resolution 
(see \cite{ChenWarren:JACM96}), using a fixed left-to-right selection
rule, for a given set of atomic (top level) queries with atoms
in $S \subseteq B_P^E$.
We will abbreviate SLG-resolution under the left-to-right selection
rule by LG-resolution.
For definite programs LG-resolution is similar to 
OLDT-resolution \cite{tamaki:sato,KanamoriKawamura:jlp93}, 
modulo the fact that OLDT specifies a more 
fixed control strategy and uses subsumption checking and term-depth
abstraction instead of variant checking.
We present a non-constructive definition of SLG-resolution that is
sufficient for our purposes, and refer to
\cite{ChenWarren:JACM96,tamaki:sato} for more
constructive formulations of (variants) of tabled resolution. 

By fixing a {\em tabling} for a program $P$,
we mean choosing a set of predicates of $P$ which are tabled.
The set of tabled predicates for a given tabling of a program $P$ is 
denoted with $Tab_P$.
The complement of this set is denoted with
$NTab_P = Pred_P \setminus Tab_P$. 

\begin{definition}[pseudo SLG-tree, pseudo LG-tree]
Let $P$ be a definite program, $Tab_P \subseteq Pred_P$, $\R$ a
selection rule and $A$ an atom.  A {\em pseudo SLG-tree w.r.t.\
$Tab_P$ for $P \cup \{ \la A\}$ under $\R$} is a tree $\tau_A$ such
that:

\begin{enumerate}
\item the nodes of $\tau_A$ are labeled with queries along with an
      indication of the selected atom according to $\R$,

\item the root of $\tau_A$ is $\la A$,

\item the children of the root $\la A$ are obtained by resolution
      against all matching program clauses in $P$, the arcs are
      labeled with the corresponding $mgu$ used in the resolution step,

\item \label{item4}
      the children of a non-root node labeled with the query ${\bf Q}$
      where $\R({\bf Q}) = B$ are obtained as follows:
      \begin{enumerate}
        \item \label{item41}
              if $Rel(B) \in Tab_P$, then

              the (possibly infinitely many) children of the node can
              only be obtained by resolving the selected atom $B$ of
              the node with clauses of the form $B\theta \la$ (not
              necessarily in $P$), the arcs are labeled with the
              corresponding $mgu$ used in the resolution step (i.e.,
              $\theta$),

        \item \label{item42}
              if $Rel(B) \in NTab_P$, then

              the children of the node are obtained by resolution of
              $B$ against all matching program clauses in $P$, and the
              arcs are labeled with the corresponding $mgu$ used in
              the resolution step.
        \end{enumerate}
\end{enumerate}
If $\R$ is the leftmost selection rule, $\tau_A$ is called a 
{\em pseudo LG-tree} w.r.t.\ $Tab_P$ for $P \cup \{ \la A\}$.
\\
We say that a pseudo SLG-tree $\tau_A$ 
w.r.t.\ $Tab_P$ for $P \cup \{ \la A\}$ is {\em smaller}
than another pseudo SLG-tree $\tau^{'}_A$ 
w.r.t.\ $Tab_P$ for $P \cup \{ \la A\}$
iff $\tau^{'}_A$ can be obtained from $\tau_A$ by attaching new sub-branches
to nodes in $\tau_A$.
\\
A {\em (computed) answer clause} of a pseudo SLG-tree $\tau_A$ 
w.r.t.\ $Tab_P$ for 
$P \cup \{ \la A\}$ is a clause of the form $A\theta \la$
where $\theta$ is the composition of the substitutions found on a branch
of $\tau_A$ whose leaf is labeled with the empty query.
\end{definition}

Intuitively, a pseudo SLG-tree (in an SLG-forest, see
Definition~\ref{defforest} below) represents the tabled computation
(w.r.t.\ $Tab_P$) of all answers for a given subquery labeling the
root node of the tree.  The trees in the above definition are called
{\em pseudo} SLG-trees because there is no condition yet on which
clauses $B\theta \la$ exactly are to be used for resolution in
point~\ref{item41}.  These clauses represent the answers found
(possibly in another tree of the forest) for the selected tabled atom.
This interaction between the trees in an SLG-forest is captured in the
following definition.

\begin{definition} 
[SLG-forest, LG-forest] \label{defforest}
Let $P$ be a definite program, 
$Tab_P \subseteq Pred_P$,
$\R$ be a selection rule and $T$ be a 
(possibly infinite) set of atoms such that no two different atoms in $T$ 
are variants of each other.
$\F$ is an {\em SLG-forest w.r.t.\ $Tab_P$
for $P$ and $T$ under $\R$} iff $\F$ is a set of 
minimal pseudo SLG-trees $\{\tau_A~|~A \in T\}$ w.r.t.\ $Tab_P$ where

\begin{enumerate}
\item $\tau_A$ is a pseudo SLG-tree w.r.t.\ $Tab_P$ for $P \cup \{\la
  A\}$ under $\R$,
\item \label{item2} every selected tabled atom $B$ of each node in
  every $\tau_A \in \F$ is a variant of an element $B^{'}$ of $T$, such
  that every clause resolved with $B$ is a variant of an answer clause
  of $\tau_{B^{'}}$ and vice versa, for
  every answer clause of $\tau_{B^{'}}$ there is a variant of this 
  answer clause which is resolved with $B$.
\end{enumerate}
Let $S$ be a set of atoms.
An {\em SLG-forest for $P$ and $S$ w.r.t.\ $Tab_P$ under $\R$} is an 
SLG-forest w.r.t.\ $Tab_P$ for a minimal set $T$ with $\tilde{S} \subseteq 
\tilde{T}$.
If $S =\{A\}$, then we also talk about the SLG-forest for $P \cup
\{\la A\}$.
\\
An {\em LG-forest} is an SLG-forest containing only pseudo LG-trees.
\end{definition}

Point \ref{item2} of Definition \ref{defforest}, together with the
imposed minimality of trees in a forest, now uniquely determines these
trees.  So we can henceforth drop the designation ``pseudo'' and refer
to (S)LG-trees in an (S)LG-forest.

Note that, selected atoms which are not tabled (i.e., of predicates
belonging to $NTab_P$) are resolved against program clauses, as in
(S)LD-resolution.  So, if $Tab_P = \emptyset $, the (S)LG-forest of $P
\cup \{\la A\}$ consists of one tree: the (S)LD-tree of $P \cup \{\la
A\}$.

We use the following small, tabled program to illustrate the notions
that we introduced so
far.  Variations of it will also be used throughout this article to
exemplify concepts related to the termination aspects of tabled logic
programs.

\begin{example} \label{exapath}
The following program $P$ computes the paths from a given node to the
reachable nodes in a given graph.  The graph is represented as a list
of terms $e(n_1,n_2)$, indicating that there is an edge from node
$n_1$ to node $n_2$;
this list is passed as an input argument to predicate
$\mathit{path}/4$ and the predicate $\mathit{edge}/3$ is used to
retrieve edges of the graph with a specific source node.
\[
\left\{
\begin{array}{lll}
path(X,Ed,Y,[Y]) & \la & edge(X,Ed,Y)
\\
path(X,Ed,Z,[Y|L]) & \la & edge(X,Ed,Y),\ path(Y,Ed,Z,L)
\\
edge(X,[e(X,Y)|L],Y) & \la & 
\\
edge(X,[e(X_1,X_2)|L],Y) & \la & edge(X,L,Y)
\end{array}
\right.
\]
Let $S = \{\mathit{path(a,[e(a,b),e(b,a)],Y,L)}\}$ and 
$Tab_P = \{\mathit{path}/4\}$.  Then,
\begin{eqnarray*}
Call(P,S) & = & \{\mathit{path(a,[e(a,b),e(b,a)],Y,L)},
                  \mathit{path(b,[e(a,b),e(b,a)],Y,L)}, \\[-2pt]
          &   & \ \mathit{edge(a,[e(a,b),e(b,a)],Y)},
                  \mathit{edge(a,[e(b,a)],Y)}, \mathit{edge(a,[~],Y)},\\[-2pt]
          &   & \ \mathit{edge(b,[e(a,b),e(b,a)],Y)},
                  \mathit{edge(b,[e(b,a)],Y)},  \mathit{edge(b,[~],Y)}\}
\end{eqnarray*}
The LG-forest w.r.t.\ $Tab_P$ for $P$ and $S$ is shown in
Figure~\ref{fig:path}.  Note that there are two LG-trees (only 2
tabled atoms are called), both with finite branches, but both trees
have an infinitely branching node.
Due to the last argument of the $\mathit{path}/4$ predicate, each of
these selected tabled atoms
has infinitely many computed answers.
\begin{figure}[hbt]
\epsfxsize=16cm
\epsfysize=11cm
\centerline{\epsffile{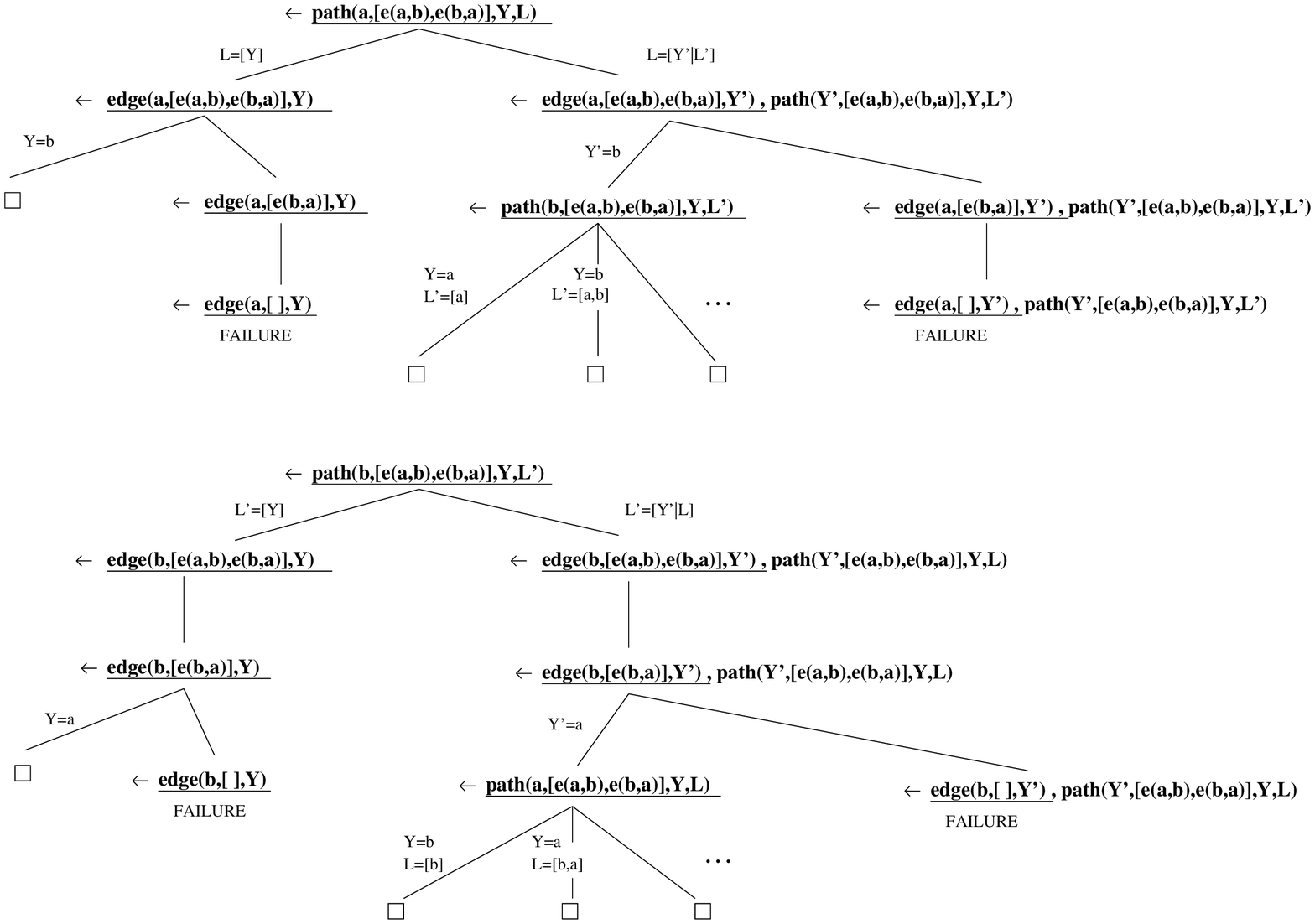}}
\caption{The LG-forest for $P \cup 
\{\la \mathit{path(a,[e(a,b),e(b,a)],Y,L)}\}$ 
w.r.t.\ $\{\mathit{path}/4\}$.}
\label{fig:path}
\end{figure}
\end{example}

As proven in e.g.\ \cite[Theorem 2.1]{KanamoriKawamura:jlp93},
the set of call patterns and the set of computed answer substitutions 
are not influenced by tabling.
Thus, we can use the notions of call set, $Call(P,S)$, and
LD-computed answer substitution, $cas$, even in the context of SLG-resolution.

The notion of a call graph (Definition~\ref{defcallgraph})
has the following particularly interesting property,
described in the proposition below,
which is useful in the study of termination.
We will use this 
property
in the proof of Theorem~\ref{theoquasichar},
which gives a necessary and sufficient condition for
quasi-termination of tabled logic programs.

\begin{proposition}[call graph: paths and selected atoms]
\label{propcallgraph}
Let $P$ be a program, $Tab_P \subseteq Pred_P$ and $S \subseteq B_P^E$.
Let $p$ be any directed path in $Call$-$Gr(P,S)$.
Then there exists an LG-derivation for some element of $Call(P,S)$,
such that all the nodes in $p$ occur as selected atoms in the
derivation.
\end{proposition}

\begin{proof}
By definition of $Call$-$Gr(P,S)$, for every arc from $\tilde{A}$ to
$\tilde{B}$ in $Call$-$Gr(P,S)$, there exists a sequence of
consecutive LG-derivation steps, starting from $\la A$ and having a
variant of $B$ as its selected atom at the end.  Because (a variant
of) $B$ is selected at the end-point, any two such derivation-step
sequences, corresponding to consecutive arcs in $Call$-$Gr(P,S)$, can
be composed to form a new sequence of LG-derivation steps.  In this
sequence, all 3 nodes of the consecutive arcs remain selected atoms in
the new sequence of derivation steps.  Transitively exploiting the
above argument yields the result.
\qed
\end{proof}

\section{Two Notions of Termination of Tabled Logic Programs}
\label{sectter}

We start by introducing a first notion of universal termination of
tabled logic programs, called \emph{quasi-termination}.  A program $P$
with a tabling $Tab_P$ is said to be quasi-terminating w.r.t.\ a query
$\la A$ iff the LG-forest of $P \cup \{\la A\}$ consists of a finite
number of LG-trees which all have finite branches.
Quasi-termination captures the property that, under LD-computation, a
given atomic query leads to only finitely many different (nonvariant)
calls to tabled predicates and there is no infinite derivation
consisting of queries with only selected non-tabled atoms.  As
mentioned in the introduction, techniques for proving
quasi-termination can be used to ensure termination of off-line
specialisation of logic programs (whether tabled or not).
Currently, in all off-line partial evaluation methods for logic programs
(e.g.~\cite{MogensenBondorf:LOPSTR92,JorgensenLeuschel:Cogen})
termination has to be ensured manually.  In the context of off-line
partial evaluation, quasi-termination (when tabling the whole set
of predicates) is actually \emph{identical} to termination of the 
partial evaluator; see e.g.\/ the discussion in~\cite{BTA@ESOP-98}.
Thus, given a technique to establish quasi-termination, one can also
establish whether a given binding time annotation will ensure
termination or whether further abstraction is called for.  This idea
has already been successfully applied in the context of functional
programming~\cite{GlenstrupJones:96:BTATermination}, using the
termination criterion of~\cite{Holst:FPCA91}.

Despite its usefulness, the notion of quasi-termination only partially
corresponds to our intuitive notion of a terminating execution of a
query against a tabled program.  This is because this notion only
requires that the LG-forest consists of only a finite number of
LG-trees, without infinite branches, yet these trees can have
infinitely branching nodes.  In order to capture this source of
non-termination for a tabled computation, we also introduce the
stronger notion of {\em LG-termination}.  A program $P$ with a tabling
$Tab_P$ is said to be LG-terminating w.r.t.\ a query $\la A$ iff the
LG-forest of $P \cup \{\la A\}$ consists of a finite number of finite
LG-trees.  So, a program $P$ is LG-terminating w.r.t.\ a query $\la A$
iff it is quasi-terminating w.r.t.\ $\la A$ and all atoms in the call
set $Call(P,\{A\})$ have only a finite number of computed answers.

In the next two subsections, we formally introduce these two notions
of termination of LG-resolution with Tabling, we give examples and
discuss some of their properties.

\subsection{Quasi-Termination} \label{ssectquasiter}

A first basic notion of universal termination under a tabled
execution mechanism is quasi-termination (a term borrowed from
\cite{Holst:FPCA91}, defining a similar notion in the context of
termination of off-line partial evaluation of functional programs).  
It is formally defined as follows.

\begin{definition}[quasi-termination]
Let $P$ be a program, $Tab_P \subseteq Pred_P$, and
$S \subseteq B_P^E$.
\\
$P$ {\em quasi-terminates w.r.t.} $Tab_P$ {\em and} 
$S$ iff
for all $A$ such that $\tilde{A} \in S$,  
the LG-forest w.r.t.\ $Tab_P$ for $P \cup \{\la A\}$
consists of a finite number of LG-trees without infinite branches.
\\
Also, $P$ {\em quasi-terminates w.r.t.} $S$ iff
$P$ quasi-terminates w.r.t.\ $Pred_P$ and $S$.
\end{definition}

Note that quasi-termination does not require that the LG-trees are
finitely branching in their nodes.

\begin{example} \label{exaquasi}
Recall the program $P$ and set $S = \{\mathit{path(a,[e(a,b),e(b,a)],Y,L)}\}$
of Example~\ref{exapath}.
The LG-forest w.r.t.\ $Tab_P = \{\mathit{path}/4\}$ was shown in
Figure~\ref{fig:path}.
$P$ quasi-terminates w.r.t.\ $Tab_P$ and $S$.
\end{example}

Many works address the problem of termination of logic programs
executed under LD-resolution (see \cite{survey} for a survey): A
program $P$ is said to be {\em LD-terminating w.r.t.} a set $S
\subseteq B_P^E$ iff for all $A$ such that $\tilde{A} \in S$, the
LD-tree of $P \cup \{\la A\}$ is finite.  In the next lemma, we show
that the notion of LD-termination is stronger than the notion of
quasi-termination.
Taking Example~\ref{exaquasi} into account,
it then follows that the notion of LD-termination is strictly stronger
than the notion of quasi-termination.

\begin{lemma} \label{lemldquasi}
Let $P$ be a program, $Tab_P \subseteq Pred_P$, and $S \subseteq B_P^E$.
\\
If $P$ LD-terminates w.r.t.\ $S$, then $P$ quasi-terminates w.r.t.\ $Tab_P$ 
and $S$.
\end{lemma}

\begin{proof}
Let $A$ be an atom such that $\tilde{A} \in S$.
Let $\F$ be the LG-forest w.r.t.\ $Tab_P$ for $P\cup \{\la A\}$.
If $P$ LD-terminates w.r.t.\ $S$, it is easy to see that $Call(P,\{A\})$ 
is finite.
Hence, $Call(P,\{A\}) \cap B_{Tab_P}^E$ is finite, and 
$\F$ consists of a finite number of LG-trees.

\noindent
Now we prove that no tree in $\F$ has an infinite branch. 
Suppose this is not the case and there is a tree in $\F$ with an infinite
branch.
Let $H$ be the leftmost atom of a query labeling a node in this infinite
branch.
Then, $H$ has an infinite LD-derivation (just plug in, for each tabled
atom $G$ in the infinite branch which is resolved with an answer,
the branch of the tree with root $G$ which leads to this answer). This
gives a contradiction.
\qed
\end{proof}

Note that by definition, $P$ quasi-terminates w.r.t.\ $Tab_P =
\emptyset$ and $S$ iff $P$ LD-terminates w.r.t.\ $S$.

Consider next the special case where all predicates occurring in $P$
are tabled.  If $Tab_P=Pred_P$, an LG-tree cannot have infinite
branches. So, $P$ quasi-terminates w.r.t.\ a set $S$ iff for all $A$
such that $\tilde{A} \in S$, the LG-forest for $P \cup \{\la A\}$
consists of a finite number of LG-trees.  The following equivalence
holds.

\begin{lemma} \label{lemspec1}
Let $P$ be a program, $Tab_P = Pred_P$, and $S \subseteq B_P^E$.
SW\\
$P$ quasi-terminates w.r.t.\ $S$ iff for all $A$ such that $\tilde{A}
\in S$, $Call(P,\{A\})$ is finite.
\end{lemma}

\begin{proof}
Since $Tab_P=Pred_P$, an LG-tree cannot have infinite branches.
The equivalence then follows from the fact that for every $A$ such
that $\tilde{A} \in S$, $B$ is the root of an LG-tree in the LG-forest
of $P \cup \{\la A\}$ iff $\tilde{B} \in Call(P,\{A\})$.
\qed 
\end{proof}

When all predicates are tabled, from the above lemma, it follows that
in case the Herbrand Universe $U_P^E$ associated to a program $P$ is
finite, $P$ quasi-terminates w.r.t.\/ any set of queries $S$.

Lemma \ref{lemspec1} does not hold in case that the tabled predicates
of a program are a strict subset of the set of predicates occurring in
the program.
A counter\-example for the if-direction is given by the program
$P =\{p \la q, q \la p\}$, the set $S= \{p\}$ and the empty set of tabled
predicates, $Tab_P = \emptyset$.
The LG-forest consists of one tree, 
namely the LD-tree of $P \cup \{\la p\}$
(so quasi-termination is the same as LD-termination).
$P$ does not quasi-terminate w.r.t.\ $Tab_P$ and $S$, whereas 
$Call(P,\{p\}) = \{p,q\}$ is a finite set.
Also the only-if direction of Lemma \ref{lemspec1}
does not hold in case $Tab_P \subset Pred_P$. 
We provide a counterexample.

\begin{example} \label{exapq}
Consider the following program $P$:
\[
\left\{
\begin{array}{lll}
p(a) & \la &
\\
p(f(X)) & \la & p(X) , q(X)
\\
q(X) & \la &
\end{array}
\right.
\]
with set of tabled predicates $Tab_P=\{p/1\}$ and $S= \{p(X)\}$.
The LG-forest is shown in Figure \ref{fig:pq}.
\begin{figure}[hbt]
\epsfysize=4.0cm
\centerline{\epsffile{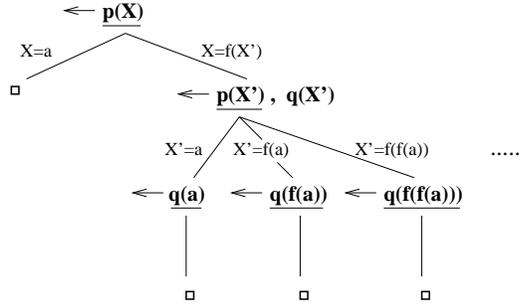}}
\caption{The LG-forest for $P \cup \{\la p(X)\}$ w.r.t.\ $\{p/1\}$.}
\label{fig:pq}
\end{figure}
$P$ quasi-terminates w.r.t.\ $Tab_P$ and $S$.  There is only one
LG-tree in the LG-forest for $P \cup \{\la p(X)\}$ without infinite
branches.  Note that the LG-tree has an infinitely branching node. 
But the call set $Call(P,\{p(X)\}) = \{p(X), q(a), \ldots,
q(f^n(a)), \ldots\}$ is infinite.
\end{example}

Note however that, 
since quasi-termination requires that there are only finitely
many LG-trees in the LG-forest of a query, there can only be a finite number
of tabled atoms in the call set of that query.
Hence, in general, the following holds.

\begin{lemma} \label{lemquasi}
Let $P$ be a program, $Tab_P \subseteq Pred_P$ and $S \subseteq B_P^E$.
\\
If $P$ quasi-terminates w.r.t.\ $Tab_P$ and $S$, then, for all $A$
such that $\tilde{A} \in S$, $Call(P,\{A\}) \cap B_{Tab_P}^E$ is
finite.
\end{lemma}

\begin{proof}
The implication follows from the fact that for every $A$ such that 
$\tilde{A} \in S$, 
$B$ is the root of an LG-tree in the LG-forest w.r.t.\ $Tab_P$ of 
$P \cup \{\la A\}$ iff $\tilde{B} \in 
(Call(P,\{A\}) \cap B_{Tab_P}^E) \cup \{\tilde{A}\}$.
\qed
\end{proof}

\begin{example}
Recall program $P$ and set $S$ of Example \ref{exapq}.
We already know that $P$ quasi-terminates w.r.t.\ $Tab_P = \{p/1\}$ and $S$.
Indeed, $Call(P,\{p(X)\}) \cap B_{Tab_P}^E = \{p(X)\}$ is finite.
\end{example}

\subsection{LG-Termination} \label{ssectlgter}

As already noted, the notion of quasi-termination only partially
corresponds to our intuitive notion of a terminating execution of a
query against a tabled program.  Therefore, the following stronger
notion of LG-termination is introduced.

\begin{definition}[LG-termination]
Let $P$ be a program, $Tab_P \subseteq Pred_P$ and $S \subseteq B_P^E$.
\\
$P$ {\em LG-terminates w.r.t.} $Tab_P$ and $S$
iff for every atom $A$ such that $\tilde{A} \in S$, 
the LG-forest w.r.t.\ $Tab_P$ for $P \cup \{\la A\}$
consists of a finite number of finite LG-trees.
\\
Also, $P$ {\em LG-terminates w.r.t.} $S$ iff $P$ LG-terminates w.r.t.\
$Pred_P$ and $S$.
\end{definition} 

Note that by definition, $P$ LG-terminates w.r.t.\ $Tab_P = \emptyset$
and $S$ iff $P$ LD-terminates w.r.t.\ $S$.

Recall the program $P$ and set $S$ of Example~\ref{exapath}.  The
LG-forest of $P$ and $S$ w.r.t.\ $Tab_P = \{\mathit{path}/4\}$ was
shown in Figure~\ref{fig:path}.  Note that there are infinitely
branching nodes in the LG-trees.  Hence, $P$ does not LG-terminate
w.r.t.\ $Tab_P$ and $S$.

Observe that if the program $P$ is called with an acyclic graph
as input, we have LG-termination and even LD-termination.
The program $P^{'}$ of the following example is obtained from $P$
by removing the 
last argument of the $\mathit{path}/4$ predicate in which the path is
computed; the resulting predicate is named $\mathit{reachable}/3$.
When $P^{'}$ is called with a cyclic graph as input, we have
LG-termination (but no LD-termination).

\begin{example}
The following program
$P^{'}$ computes the reachable nodes 
from a given node in a given graph.
As in Example~\ref{exapath},
the graph is represented as a list of terms $e(n_1,n_2)$, indicating that
there is an edge from node $n_1$ to node $n_2$.
Note that, 
contrary to
program $P$ of Example~\ref{exapath},
$P^{'}$ does not compute the paths leading from the given node
to the reachable nodes.
\[
\left\{
\begin{array}{lll}
reachable(X,Ed,Y) & \la & edge(X,Ed,Y)
\\
reachable(X,Ed,Z) & \la & edge(X,Ed,Y),\ reachable(Y,Ed,Z)
\\
edge(X,[e(X,Y)|L],Y) & \la & 
\\
edge(X,[e(X_1,X_2)|L],Y) & \la & edge(X,L,Y)
\end{array}
\right.
\]
Let $S^{'} = \{\mathit{reachable(a,[e(a,b),e(b,a)],Y)}\}$ and
$Tab_{P^{'}} = \{\mathit{reachable}/3\}$..
Then,
\begin{eqnarray*}
Call(P^{'},S^{'}) & = & \{\mathit{reachable(a,[e(a,b),e(b,a)],Y)},
                          \mathit{reachable(b,[e(a,b),e(b,a)],Y)},\\[-2pt]
                  &   & \ \mathit{edge(a,[e(a,b),e(b,a)],Y)},
                          \mathit{edge(a,[e(b,a)],Y)},
                          \mathit{edge(a,[~],Y)},\\[-2pt]
                  &   & \ \mathit{edge(b,[e(a,b),e(b,a)],Y)},
                          \mathit{edge(b,[e(b,a)],Y)},
                          \mathit{edge(b,[~],Y)}\}
\end{eqnarray*}
The LG-forest w.r.t.\ $Tab_{P^{'}}$ for $P^{'}$ and $S^{'}$ is shown
in Figure~\ref{fig:reach}.  Note that there are 2 LG-trees (only 2
tabled atoms are called), both with finite branches and finitely
branching nodes (the selected tabled atoms have a finite number of
computed answers).  $P^{'}$ LG-terminates w.r.t.\ $Tab_{P^{'}}$ and
$S^{'}$.  Observe that $P^{'}$ does not LD-terminate w.r.t.\ $S^{'}$.
\begin{figure}[hbt]
\epsfxsize=16cm
\epsfysize=11.0cm
\centerline{\epsffile{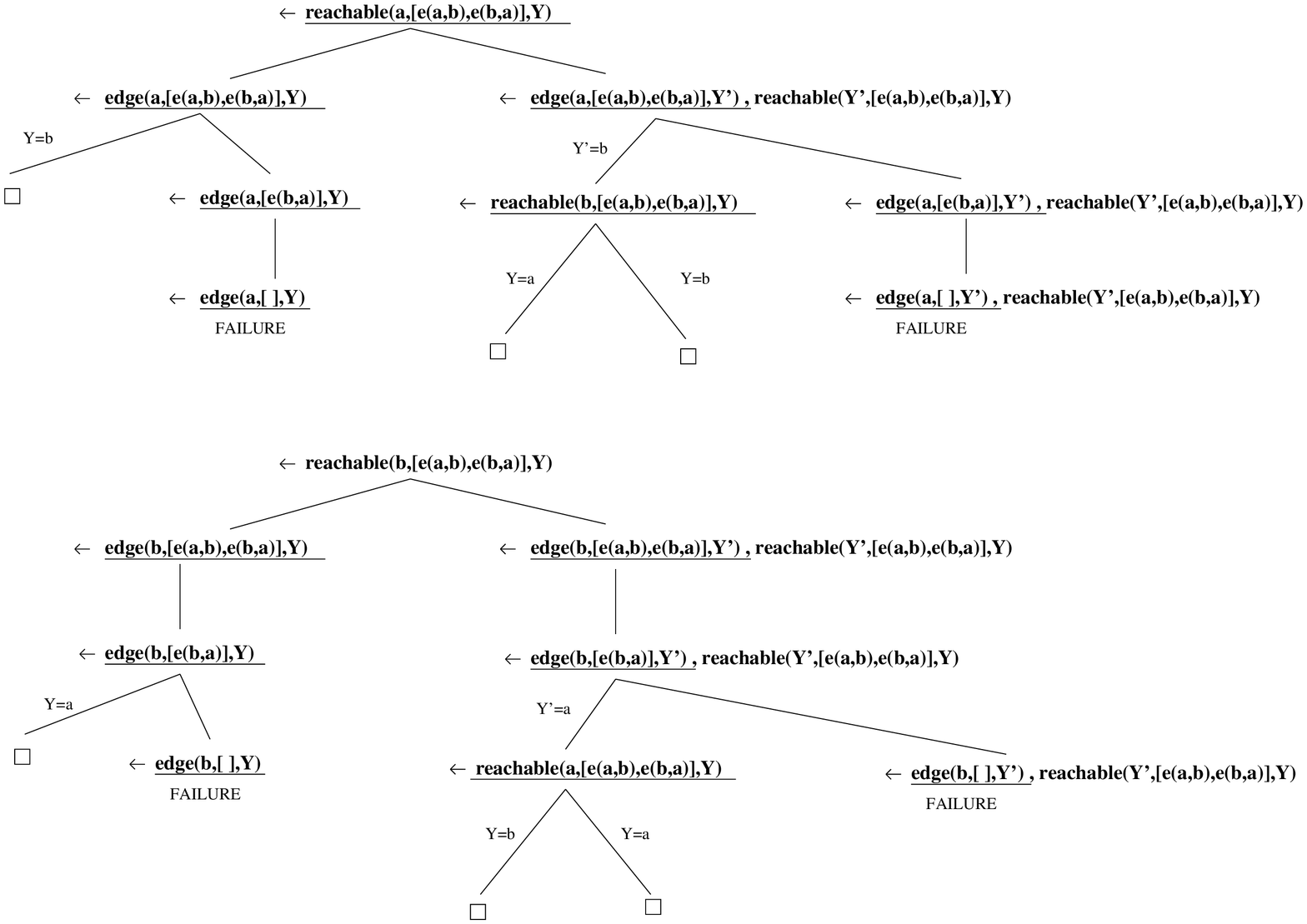}}
\caption{The LG-forest for $P^{'} \cup 
\{\la \mathit{reachable(a,[e(a,b),e(b,a)],Y)}\}$ w.r.t.\ 
$\{\mathit{reachable}/3\}$.}
\label{fig:reach}
\end{figure}
\end{example}

As illustrated by the above examples, the notion of LG-termination is
strictly stronger than the notion of quasi-termination.  Also,
LD-termination implies (and is strictly stronger than) LG-termination.

\begin{lemma} \label{lemrel}
Let $P$ be a program, $Tab_P \subseteq Pred_P$, and $S \subseteq B_P^E$.
\\
If $P$ LG-terminates w.r.t.\ $Tab_P$ and $S$, 
then $P$ quasi-terminates w.r.t.\ $Tab_P$ and $S$.
\\
If $P$ LD-terminates w.r.t.\ $S$,
then $P$ LG-terminates w.r.t.\ $Tab_P$ and $S$.
\end{lemma}

\begin{proof}
The first statement is trivial by definition.
For the second statement, this is a corollary of the following
Proposition \ref{proprel} with
$Tab_1 = \emptyset$ and $Tab_2 = Tab_P$.
\qed \end{proof}

Note that, if a program quasi-terminates w.r.t.\ a tabling and a set
$S$ and the program does not LG-terminate w.r.t.\ that tabling and $S$,
then there does not exist a tabling such that the program LG-terminates
w.r.t.\ that tabling and $S$. This is proven in the following lemma.

\begin{lemma}
Let $P$ be a program and $S\subseteq B_P^E$ a set of queries.
Suppose there exists a tabling $Tab^{\ast}_P \subseteq Pred_P$ such that 
$P$ quasi-terminates w.r.t.\ $Tab^{\ast}_P$ and $S$ and $P$ does not
LG-terminate w.r.t.\ $Tab^{\ast}_P$ and $S$.
\\
Then for all tablings $Tab_P \subseteq Pred_P$,
$P$ does not LG-terminate w.r.t.\ $Tab_P$ and $S$.
\end{lemma}

\begin{proof}
Let $Tab^{\ast}_P \subseteq Pred_P$ be such that 
$P$ quasi-terminates w.r.t.\ $Tab^{\ast}_P$ and $S$ and $P$ does not
LG-terminate w.r.t.\ $Tab^{\ast}_P$ and $S$.
Then, there exists a predicate $p \in Tab^{\ast}_P \cap Rec_P$
such that there is a $p$-atom in $Call(P,S)$ which has infinitely
may different (nonvariant) computed answers.
Since tabling does not influence 
the set of call patterns nor the set of 
computed answer substitutions (see e.g.
\cite[Theorem 2.1]{KanamoriKawamura:jlp93}),
there cannot exist a tabling such that $P$ LG-terminates
w.r.t.\ that tabling and the set $S$.
\qed
\end{proof}

Consider two tablings $Tab_1, Tab_2 \subseteq Pred_P$ for a program $P$.
Suppose $Tab_1 \subseteq Tab_2$ (hence $NTab_1 \supseteq NTab_2$).
The next proposition studies the relationship 
between the LG-termination of $P$ w.r.t.\ these two tablings.

\begin{proposition} \label{proprel}
Let $P$ be a program.
Let $Pred_P = Tab_1 \sqcup NTab_1$ and $Pred_P = Tab_2 \sqcup NTab_2$.
Suppose $Tab_1 \subseteq Tab_2$.
Let $S \subseteq B_P^E$.
\\
If $P$ $LG$-terminates w.r.t.\ $Tab_1$ and $S$, then
$P$ $LG$-terminates w.r.t.\ $Tab_2$ and $S$.
\end{proposition}

\begin{proof}
Let $A$ be an atom such that $\tilde{A} \in S$.
Let $\F_1$ be the LG-forest w.r.t.\ $Tab_1$ of $P \cup \{\la A\}$ and 
let $\F_2$ be the LG-forest w.r.t.\ $Tab_2$ of $P \cup \{\la A\}$.
We know that $\F_1$ consists of a finite number of finite LG-trees.
So, $\sharp Call(P,\{A\}) < \infty$, hence, 
$\sharp( Call(P,\{A\}) \cap B_{Tab_2}^E) < \infty$ and $\F_2$ consists
of a finite number of LG-trees.
We prove that the LG-trees of $\F_2$ are finite.
Since each LG-tree in $\F_2$ can be extended to obtain an LG-tree in $\F_1$, 
this follows from the finiteness of the LG-trees in $\F_1$. 
\qed \end{proof}

Note that this proposition does not hold for quasi-termination as 
is shown in the following example.

\begin{example}
Recall the program $P$ and set $S =\{p(X)\}$ of Example \ref{exapq}.
Let $Tab_1 = \{p/1\}$ (as in Example \ref{exapq}) 
and $Tab_2 = \{p/1, q/1\}$.
Then, $P$ quasi-terminates w.r.t.\ $Tab_1$ and $S$
(the LG-forest in this case was shown in Figure \ref{fig:pq}).
But, as is shown in Figure \ref{fig:pq2},
$P$ does not quasi-terminate w.r.t.\ $Tab_2$ and $S$.
\begin{figure}[hbt]
\epsfxsize=12.0cm
\epsfysize=4.0cm
\centerline{\epsffile{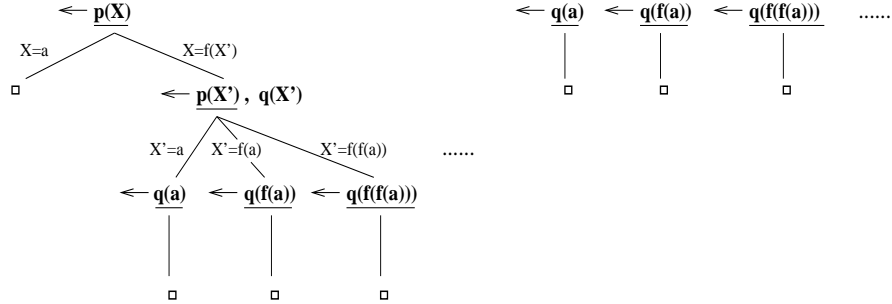}}
\caption{The LG-forest for $P \cup \{\la p(X)\}$ w.r.t.\ $\{p/1,q/1\}$.}
\label{fig:pq2}
\end{figure}
\end{example}

\subsection{Characterization of LG-termination through quasi-termination}

We now relate the notions of quasi-termination and LG-termination in a
more detailed way.  By definition, quasi-termination only corresponds
to part of the LG-termination notion; it fails to capture
non-termination caused by an infinitely branching node in an LG-tree.
Note that if an LG-forest contains a tree with an infinitely branching
node, then there is an LG-tree in the forest which is infinitely
branching in a node which contains a query with a selected atom which
is tabled and recursive.  This observation leads to the following
lemma.  We denote the set of tabled, recursive predicates in a program
$P$ with $TR_P$:
\[
        TR_P = Tab_P \cap Rec_P .
\]

\begin{lemma} \label{lemlgq}
Let $P$ be a program, $Tab_P \subseteq Pred_P$, and $S \subseteq B_P^E$.
\\
$P$ LG-terminates w.r.t.\ $Tab_P$ and $S$
iff
$P$ quasi-terminates w.r.t.\ $Tab_P$ and $S$ and
for all $A$ such that $\tilde{A} \in S$, the set of LD-computed answers for
atoms in $Call(P,\{A\}) \cap B^E_{TR_P}$ is finite.
\end{lemma} 

\begin{proof}
$\Rightarrow :$
Suppose $P$ LG-terminates w.r.t.\ $Tab_P$ and $S$. 
Then $P$ quasi-terminates w.r.t.\ $Tab_P$ and $S$.
It is easy to see that, since for every $A$ such that
$\tilde{A} \in S$
the LG-forest for $P\cup \{\la A\}$ consists of a finite
number of finite trees, the set
of computed answers for atoms in $Call(P,\{A\})$ is finite.

\noindent
$\Leftarrow :$ Suppose that
$P$ quasi-terminates w.r.t.\ $Tab_P$ and $S$ and
for all $A$ such that $\tilde{A} \in S$ the set of LD-computed answers for
atoms in $Call(P,\{A\}) \cap B^E_{TR_P}$ is finite.
We prove that $P$ LG-terminates w.r.t.\ $Tab_P$ and $S$.
Let $A$ be an atom such that $\tilde{A} \in S$.
We already know that the LG-forest $\F$ of $P\cup \{\la A\}$ consists
of a finite number of LG-trees without infinite branches.
We prove by contradiction that these LG-trees are finitely branching.
Suppose there is an LG-tree in $\F$ which is infinitely branching.
Then, there is an LG-tree in $\F$ with an infinitely branching node, 
which contains a query which
has a tabled, recursive atom at the leftmost position.
That is, there is an atom in $Call(P,\{A\}) \cap B^E_{TR_P}$
which has infinitely many computed answers. 
This gives a contradiction.
\qed \end{proof}

It follows from the proof of Lemma \ref{lemlgq} that, 
if $P$ LG-terminates w.r.t.\ $Tab_P$ and $S$,
the set of computed answers for atoms in $Call(P,\{A\})$ is finite
for all $A$ such that $\tilde{A} \in S$.

Based on the observation in Lemma~\ref{lemlgq}, we next define a
transformation on programs, called the answer-transformation, such
that LG-termination of a program $P$ is equivalent to the
quasi-termination of the program $P^a$ obtained by applying the
answer-transformation on $P$.

\begin{definition}
[a(nswer)-transformation] \label{defgsol}
Let $P$ be a program and $Tab_P \subseteq Pred_P$.
The {\em a-transformation} on $P$ and $Tab_P$ is defined as follows:

\begin{itemize}

\item

For a clause $C = H \la B_1, \ldots, B_n$ in $P$, we define
\[
C^{a} = H \la B_1, B_1^*, \ldots, B_n, B_n^*
\]
with $B_i^*$ defined as follows (suppose $B_i =p(t_1,\ldots,t_n)$):

if $p \in Tab_P$ and $p \simeq Rel(H)$
then
$B_i^* = p^a(t_1,\ldots,t_n)$, where $p^a/n$ is a new predicate,
else
$B_i^* = \emptyset$.

Let $TR^{a}_P = \{p^a/n~|~p/n \in TR_P \}$
(recall that $TR_P = Tab_P \cap Rec_P$).

\item
For the program $P$, we define
$$
        P^{a} = \{C^{a}~|~C \in P\}
                \cup
                \{p^a(X_1, \ldots, X_n) \la~|~p^a/n \in TR^a_P\} .
$$

\item
The set of tabled predicates of the program $P^{a}$ 
is defined as  
\[
        Tab_{P^{a}} = Tab_P \cup TR^a_P.
\]
\end{itemize}

\end{definition}

\begin{example} \label{exaatransf}
Let $P$ be the program of Example~\ref{exapath}, with 
$Tab_P = \{\mathit{path}/4\}$.
The a-transformation, $P^a$, of $P$ is the following program:
\[
\left\{
\begin{array}{lll}
path(X,Ed,Y,[Y]) & \la & edge(X,Ed,Y)
\\
path(X,Ed,Z,[Y|L]) & \la & edge(X,Ed,Y),\ path(Y,Ed,Z,L),\ 
\\
 & & path^a(Y,Ed,Z,L)
\\
edge(X,[e(X,Y)|L],Y) & \la & 
\\
edge(X,[e(X_1,X_2)|L],Y) & \la & edge(X,L,Y)
\\
path^a(X,Ed,Y,L) & \la &
\end{array}
\right.
\]
with
$Tab_{P^a} = \{\mathit{path}/4, \mathit{path^a}/4\}$.
\end{example}

It is easy to see that $Call(P,S) = Call(P^a,S) \cap B_P^E$.
Also, if we denote with $cas(P,\{p(\overline{t})\})$ the set of 
computed answer substitutions for $p(\overline{t})$ in $P$,
then $cas(P,\{p(\overline{t})\})$ $= cas(P^a,\{p(\overline{t})\})$
for all $p(\overline{t}) \in B_P^E$.
It is important to note that, if we have a query $p(\overline{t})
\in B^E_{TR_P}$ to the program $P$, then
$p(\overline{t})\sigma$ is a computed answer if 
$p^a(\overline{t})\sigma$ $\in Call(P^a,\{p(\overline{t})\})$.
This is in fact the main purpose of the transformation.

\begin{theorem} 
[characterisation of LG-termination in terms of quasi-ter\-mination]
\label{theolg}
Let $P$ be a program, $Tab_P \subseteq Pred_P$
and $S \subseteq B_P^E$.
\\
$P$ LG-terminates w.r.t.\ $Tab_P$ and $S$ iff
$P^{a}$ quasi-terminates w.r.t.\ 
$Tab_{P^{a}}$ and $S$.
\end{theorem}

\begin{proof}
$\Leftarrow:$
Suppose $P^a$ is quasi-terminating 
w.r.t.\ $Tab_{P^a}$ and $S$.
Let $A$ be an atom such that $\tilde A \in S$. Let $\F$ be the LG-forest 
w.r.t.\ $Tab_P$ of 
$P\cup \{\la A\}$. We prove that $\F$ consists of a finite number of finite 
LG-trees.
\\
We know that the LG-forest $\F^a$ w.r.t.\
$Tab_{P^a}$ of $P^a\cup \{\la A\}$
is a finite set of LG-trees, without infinite branches.
It is easy to see that hence, $\F$ consists also of a finite 
number of trees without infinite branches.
We prove that 
the LG-trees in $\F$ are finitely branching.
Suppose this is not the case, i.e.\ there is an LG-tree in $\F$ which is 
infinitely branching.
Then, there is an LG-tree $T$ in $\F$ which is infinitely branching 
in a non-root
node, which is a query with leftmost atom 
$p(t_1,\ldots,t_n)$, with $p \in TR_P$, which is directly descending
from an atom $q(s_1,\ldots,s_m)$, with $p \simeq q$, via a 
recursive clause $C =q(u_1,\ldots,u_m) \la \ldots, p(v_1,\ldots,v_n), \ldots$.
Let $T^a$ be the LG-tree in $\F^a$ corresponding to $T$.
Note that the clause $C^a$ instead of $C$ is used in $T^a$.
Because of this, the atom to the right of 
$p(t_1,\ldots,t_n)$ in 
the infinitely branching node 
is $p^a(t_1,\ldots,t_n)$. Thus, 
$\F^a$ consists of a infinite number of LG-trees (there are an infinite number
of LG-trees with predicate $p^a$ in the root).
This gives a contradiction.
\\
$\Rightarrow:$
Suppose $P$ is LG-terminating w.r.t.\ $Tab_P$ and $S$.
Let $A$ be an atom such that $\tilde A \in S$. Let $\F$ be the LG-forest 
w.r.t.\ $Tab_P$ of $P\cup \{\la A\}$.
Then, $\F$ consists of a finite number of finite LG-trees.
Let $\F^a$ be the LG-forest 
w.r.t.\ $Tab_{P^a}$ of $P^a\cup \{\la A\}$.
By definition of the $a$-transformation, we see that 
$\F^a$ also consists of a finite number of finite LG-trees.
Hence, $P^a$ quasi-terminates (and even LG-terminates) w.r.t.\ $Tab_{P^a}$
and $S$.
\qed \end{proof}

\begin{example}[Example~\ref{exaatransf} continued]
The LG-forest w.r.t.\ $Tab_P$
of $P$ and 
\\
$\{path(\mathit{a,[e(a,b),e(b,a)],Y,L})\}$ 
was shown in Figure~\ref{fig:path}.
Note that the trees are infinitely branching
and hence $P$ does not LG-terminate w.r.t.\ $Tab_{P}$ and
$\{\mathit{path(a,[e(a,b),e(b,a)],Y,L)}\}$.
\\
In Figure~\ref{fig:patha}, the LG-forest of the program $P^a$ and 
$\{\mathit{path(a,[e(a,b),e(b,a)],Y,L)}\}$ w.r.t.\ $Tab_{P^a}$ is shown.
Note that there are infinitely many LG-trees in the forest; 
$P^a$ does not quasi-terminate w.r.t.\ $Tab_{P^a}$ and
$\{\mathit{path(a,[e(a,b),e(b,a)],Y,L)}\}$.

\begin{figure}[p]
\epsfxsize=17cm
\epsfysize=18cm
\centerline{\epsffile{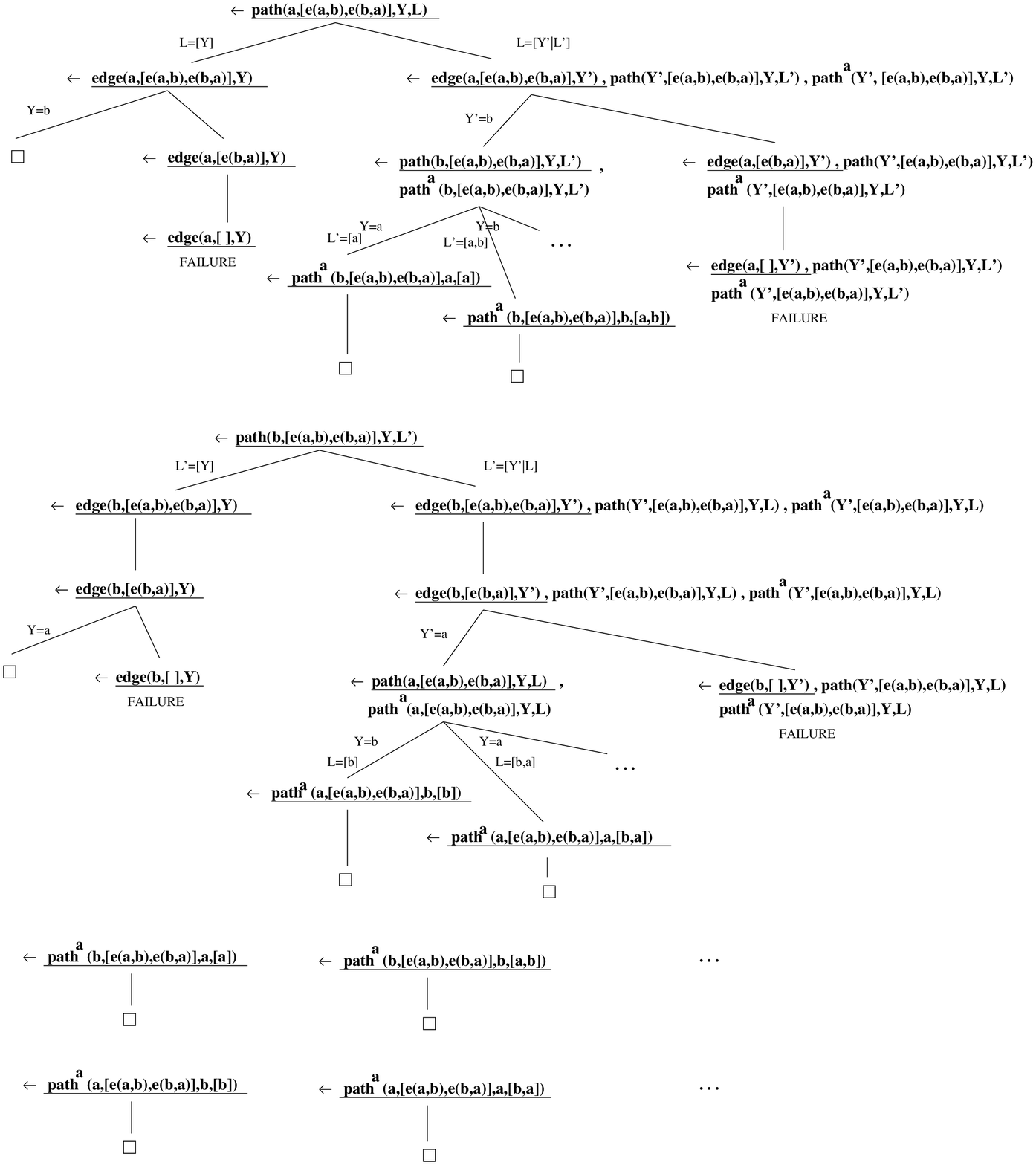}}
\caption{The LG-forest for $P^a \cup
\{\la path(\mathit{a,[e(a,b),e(b,a)],Y,L})\}$ w.r.t.\/ 
$Tab_{P^a}$.}
\label{fig:patha}
\end{figure}

\end{example}

\section{Conditions for Termination of Tabled Logic Programs}
\label{sectchar}

In this section, we give sufficient conditions for the notions of
quasi-termination and LG-termination.  We prove that these conditions
are also necessary in case the tabling satisfies the property of being
{\em well-chosen}.  First, we want to note that the termination
conditions are adapted from the acceptability notion for
LD-termination defined in \cite{framework}, and not from the more
``standard'' definition of acceptability by Apt and Pedreschi 
in~\cite{apt:pedreschi}.  The reason for this choice is that the
quasi-termination as well as the LG-termination property of a tabled
program and query is \emph{not}~closed under substitution.  To see
this, consider the following example from
\cite{LeuschelMartensSagonas}.

\begin{example} \label{exanotsubstclosed}
Let $p/2$ be a tabled predicate defined by the following clause.
\[
p(f(X),Y) \la p(X,Y)
\]
Then, the query $\la p(X,Y)$ terminates while $\la p(X,X)$ does not.
\end{example}

The acceptability notion in \cite{apt:pedreschi} is expressed in terms
of ground instances of clauses and its associated notion of
LD-termination is expressed in terms of the set of all queries that
are bounded under the given level mapping. Such sets are closed under
substitution. Because quasi-termination lacks invariance under
substitution, we need a stronger notion of acceptability, capable of
treating \emph{any} set of queries.

We next introduce the notion of well-chosen tabling w.r.t.\ a program.
If the tabling is well-chosen, we are able to give a necessary and
sufficient condition for quasi-termination and for LG-termination.
If the tabling is not well-chosen, the condition is still sufficient.

We first introduce some notation.
Let $P$ be a program and let $G_P$ be the dependency graph 
of the predicates of $P$.
For a tabling $Tab_P$ for $P$ and predicates $p, q
\in NTab_P$ with $p \simeq q$, let $C_1(p,q),$ $C_2(p,q)$ and $C_3(p,q)$
denote the following disjoint cases:
\begin{description}
\item[$C_1(p,q)$:] No cycle of directed arcs in $G_P$ containing $p$
  and $q$ contains a predicate from $Tab_P$.
\item[$C_2(p,q)$:] All cycles of directed arcs in $G_P$ containing $p$
  and $q$ contain at least one predicate from $Tab_P$.
\item[$C_3(p,q)$:] There is a cycle of directed arcs in $G_P$ containing $p$
  and $q$ which contains no predicate from $Tab_P$ and there is a 
  cycle of directed arcs in $G_P$ containing $p$
  and $q$ which contains a predicate from $Tab_P$.
\end{description}
Note that $C_1(p,q),$ $C_2(p,q)$ and $C_3(p,q)$ depend on the
program $P$ (more precisely on the dependency graph $G_P$ of $P$)
and on the tabling $Tab_P$ for $P$. 
When referring to one of these three cases, it will always be clear
from the context which program and tabling are under consideration.
Given a program $P$ and tabling $Tab_P$, for all predicates $p, q \in
NTab_P$ with $p \simeq q$, exactly one of the cases $C_1(p,q),$
$C_2(p,q)$ or $C_3(p,q)$ holds.

\begin{example} \label{ex:wellchosen1}
Consider the following three propositional programs $P$, $P^{'}$ and
$P^{''}$:
\[
\begin{array}{lllll}
P:
\left\{
\begin{array}{lll}
a & \la & b \\
b & \la & c \\
c & \la & b
\end{array}
\right.
& \qquad &
P^{'}: 
\left\{
\begin{array}{lll}
a & \la & b \\
b & \la & c \\
c & \la & a
\end{array}
\right.
& \qquad &
P^{''}: 
\left\{
\begin{array}{lll}
a & \la & b \\
b & \la & c \\
c & \la & a \\
c & \la & b
\end{array}
\right.
\end{array}
\]
with $Tab_{P} = Tab_{P^{'}} = Tab_{P^{''}} = \{a/0\}$.

\noindent
For the program $P$, we have that $C_1(b,c)$ holds.
For the program $P^{'}$, we have that $C_2(b,c)$ holds.
For the program $P^{''}$, we have that $C_3(b,c)$ holds.
\end{example}

We next define the notion of well-chosen tabling w.r.t.\ a program
$P$.  A tabling for $P$ is well-chosen w.r.t.\ $P$ if it is such that
the third case $C_3$ never occurs.

\begin{definition} 
[well-chosen tabling (w.r.t.\ a program)] \label{assump}
Let $P$ be a program.
The tabling $Tab_P$ is called {\em well-chosen} w.r.t.\ the program
$P$ iff for every $p, q \in NTab_P$ such that 
$p \simeq q$, either $C_1(p,q)$ or $C_2(p,q)$ holds.
\end{definition}

Note that in case $Tab_P$ is well-chosen w.r.t.\ $P$, 
we have that if $p, q, r \in NTab_P$ and
$p \simeq q \simeq r$ and $C_1(p,q)$ (resp.\ $C_2(p,q)$) holds, then
$C_1(q,r)$ (resp.\/ $C_2(q,r)$) holds.
In the special case that $NTab_P \subseteq \{ p \in Pred_P~|~p$
is a non-recursive or only directly recursive predicate$\}$
or that $NTab_P = \emptyset$ (i.e.\/ $Tab_P = Pred_P$),
the tabling $Tab_P$ is well-chosen w.r.t.\/ $P$.

\begin{example} \label{ex:wellchosen2}
Recall the programs $P$, $P^{'}$ and $P^{''}$ of Example \ref{ex:wellchosen1}.
The tabling $\{a/0\}$ is well-chosen w.r.t.\/ $P$ and $P^{'}$, but
not w.r.t.\/ $P^{''}$.
\end{example}

\subsection{Quasi-Termination} \label{ssectquasichar}

We now introduce the notion of quasi-acceptability, in general a
sufficient condition for quasi-termination.  In case the tabling is
well-chosen, quasi-acceptability is also a necessary condition for
quasi-termination.

\begin{definition}[quasi-acceptability] \label{defquasiacce}
Let $P$ be a program, $Tab_P \subseteq Pred_P$,
and $S \subseteq B_P^E$.
$P$ is {\em quasi-acceptable} w.r.t.\ $Tab_P$ and $S$ iff
there is a level mapping $\level{.}$ on $B_P^E$ such that
for all $A$ such that $\tilde{A} \in S$, $\level{.}$ 
is finitely partitioning on $Call(P,\{A\}) \cap B_{Tab_P}^E$ and
such that
\begin{itemize}

\item
for every atom $A$ such that $\tilde{A} \in Call(P,S)$,

\item
for every clause $H \la B_1, \ldots, B_n$ in $P$, such that $mgu(A,H)= \theta$
exists,

\item
for every $1 \leq i \leq n$,

\item
for every $cas$ 
$\theta_{i-1}$ for $\la (B_1, \ldots, B_{i-1})\theta$:
\[
\begin{array}{ll}
\level{A} \geq  \level{B_i \theta \theta_{i-1}} &
\\
\mbox{and} &
\\
\level{A} >  \level{B_i \theta \theta_{i-1}} & \mbox{if}~Rel(A) 
\simeq Rel(B_i) \in NTab_P~\mbox{and}
\\
 & C_2(Rel(A),Rel(B_i))~\mbox{does not hold} .
\end{array}
\]
\end{itemize}
\end{definition}

\begin{theorem}[(necessary and) sufficient condition for 
quasi-termination] \label{theoquasichar}
Let $P$ be a program, $Tab_P \subseteq Pred_P$
and $S \subseteq B_P^E$.
\\
If $P$ is quasi-acceptable w.r.t.\ $Tab_P$ and $S$,
then $P$ quasi-terminates w.r.t.\ $Tab_P$ and $S$.
\\
If the tabling $Tab_P$ is well-chosen w.r.t.\ $P$, then also the
converse holds, i.e.\ $P$ is quasi-acceptable w.r.t.\ $Tab_P$ and $S$
iff $P$ quasi-terminates w.r.t.\ $Tab_P$ and $S$.
\end{theorem}

\begin{proof}

$\Rightarrow :$ Suppose that $P$ is quasi-acceptable w.r.t.\ $Tab_P$,
$S$ and a level mapping $\level{.}$.  We prove that $P$
quasi-terminates w.r.t.\ $Tab_P$ and $S$.  Let $A$ be an atom such
that $\tilde{A} \in S$, let $\F$ be the LG-forest w.r.t.\ $Tab_P$ of
$P \cup \{\la A\}$.

\begin{itemize}

\item
$\F$ consists of a finite number of LG-trees, i.e.\
$\sharp( Call(P,\{A\}) \cap B_{Tab_P}^E ) < \infty$.

Due to the quasi-acceptability condition, 
any call in $Call(P,\{A\})$ directly
descending from $A$, say $B$, is such that $\level{A} \geq \level{B}$.
The same holds recursively for the atoms descending from $B$.
Thus, the level mapping of any call, recursively descending from $A$, 
is smaller than or equal to $\level{A} \in \bbbn$.
Since $\level{.}$ is finitely partitioning on $Call(P,\{A\}) \cap
B_{Tab_P}^E$, we have that: 
$\sharp( \bigcup_{n \leq \level{A}}\level{.}^{-1}(n) \cap Call(P,\{A\}) \cap
B_{Tab_P}^E) < \infty$.
Hence, 
$\sharp( Call(P,\{A\}) \cap B_{Tab_P}^E ) < \infty$, i.e.\
$\F$ consists of a finite number of trees. 

\item
The LG-trees in $\F$ have finite branches.

Suppose there is a tree in $\F$ with an infinite branch. This infinite branch
contains an infinite directed subsequence
$G_0, G_1, \ldots$. 
It is easy to see that the leftmost atoms in the nodes 
of this infinite directed 
subsequence all are $NTab_P$-atoms (because $Tab_P$-atoms are resolved using
answers).
There is a $n \in \bbbn$, such that 
each $G_i$, $i \geq n$, has as leftmost atom $A_i$ and 
for all $i \geq n$, $Rel(A_i) \simeq Rel(A_{i+1})$ and 
$C_2(Rel(A_i),Rel(A_{i+1}))$ does not hold.
Because of the quasi-acceptability condition,
$\level{A_i} > \level{A_{i+1}}$, for all $i \geq n$.
This gives a contradiction.
\end{itemize}

\noindent
$\Leftarrow :$
Suppose that the tabling $Tab_P$ is well-chosen w.r.t.\ $P$ and suppose
that $P$ quasi-terminates w.r.t.\ $S$. 
We have to construct 
a level mapping $\level{.}$ such that 
$P$ is quasi-acceptable w.r.t.\ $Tab_P$, $S$ and this level mapping
$\level{.}$.
We will only define $\level{.}$ on elements of $Call(P,S)$. On elements of the 
complement of $Call(P,S)$ in $B_P^E$, $\level{.}$ can be assigned any value,
as these elements do not turn up in the quasi-acceptability condition.
\\
In order to define $\level{.}$ on $Call(P,S)$, 
consider the $Call$-$Gr(P,S)$-graph
(Definition \ref{defcallgraph}). 
Consider a strongly connected component $C$ in $Call$-$Gr(P,S)$.
\\
Then, there is at least one $Tab_P$-atom in $C$.
To see this, suppose this is not the case. Consider a cyclic path $p$ in $C$.
This consists only of $NTab_P$-atoms. But then, because of Proposition 
\ref{propcallgraph}, there is an infinite
branch in a tree of the LG-forest of an element of $S$. This gives a 
contradiction.
\\
Also, there is only a finite number of $Tab_P$-atoms in $C$.
To see this, suppose this is not the case.
Then there is an infinitely long path $p$ through infinitely many $Tab_P$-atoms
of $C$.  Because of Proposition \ref{propcallgraph},
there is an infinite number of $Tab_P$-atoms selected in a derivation of an 
element of $S$, i.e.\/ there are infinitely many trees in the LG-forest 
of that element of $S$.  This gives a contradiction.
\\
For every two non-tabled atoms, say $p(\overline{t})$ and
$q(\overline{s})$, in $C$ (note that thus $p \simeq q$), $C_1(p,q)$
does not hold (since there is at least one $Tab_P$-atom in $C$).
Thus, since the tabling is well-chosen, $C_2(p,q)$ holds.
\\
Define $\overline{CG}$ as the graph obtained from $Call$-$Gr(P,S)$
by replacing any strongly connected component by a single contracting
node and replacing any arc from $Call$-$Gr(P,S)$ pointing to (resp.\ from)
any node in that strongly connected component by an arc to (resp.\ from)
that contracting node.
$\overline{CG}$ does not have any (non-trivial) strongly connected
components. Moreover, any strongly connected component from $Call$-$Gr(P,S)$
that was collapsed into a contracting node of 
$\overline{CG}$ necessarily contains at least one
and at most a finite number of $Tab_P$-atoms.
\\
Note now that each path in $\overline{CG}$ which is not cyclic 
(there are only trivial cycles in $\overline{CG}$) is 
finite. This also follows directly from Proposition
\ref{propcallgraph}.
\\
Note also that it is possible that $\overline{CG}$ has an infinitely
branching (possibly contracting) node.
Let $A$ be an atom in that infinitely branching node.
It follows from Lemma \ref{lemquasi} 
that, because $P$ quasi-terminates w.r.t.\ $S$, 
$\sharp(\{B~|~B$ is a descendant of $A$ in 
$\overline{CG}\}$ $\cap$ $B^E_{Tab_P}) <$ $\infty$.
\\
We now construct $\overline{\overline{CG}}$ from 
$\overline{CG}$ 
starting from the top nodes $N_1$ downwards as follows:
\begin{itemize}
\item replace all direct descendants of $N_1$ in $\overline{CG}$
      different from $N_1$, by a single contracting node $N_2$;

\item replace any arc from $\overline{CG}$ pointing to (resp.\ from)
      any node in that (possibly infinite) set of direct descendants
      by an arc to (resp.\ from) that contracting node $N_2$;

\item repeat this for the nodes $N_2$.
\end{itemize}

\noindent
This process stops because, as we already noted, 
each path in $\overline{CG}$ which is not cyclic is finite.
It is easy to see that $\overline{\overline{CG}}$ is a graph in which
each node has at most one direct descendant different from itself.
Also, each node in $\overline{\overline{CG}}$ consists of a 
(possibly infinite) set of nodes of $Call$-$Graph(P,S)$ which
contains only finitely many $Tab_P$-atoms.
\\
We define the level mapping $\level{.}$ as follows.
Consider the layers of $\overline{\overline{CG}}$ (there are 
only a finite number of layers).
Let layer-0 be the set of leaves in $\overline{\overline{CG}}$.
We assign to these nodes a number in $\bbbn$, such that
all nodes get a different number.
Then, we move up to the next layer in $\overline{\overline{CG}}$.
This layer, layer-1, consists of all nodes $N$ such that
the path starting from $N$ has length 1.
We assign to each such node $N$ a natural number, 
such that the number assigned to $N$ is strictly larger than the number
assigned to its descendant (in the previous step).
We continue this process layer by layer.
The value of the level mapping $\level{.}$ 
on elements of $Call(P,S)$ is defined as 
follows:
all calls contained in the node $N$ receive the 
number assigned to the node $N$.
\\
We prove that $P$ is quasi-acceptable w.r.t.\ $Tab_P$, $S$ and
this level mapping $\level{.}$. 
\begin{itemize}

\item
for every $A \in S$,
$\level{.}$ is finitely partitioning on $B_{Tab_P}^E \cap Call(P,\{A\})$.

Note that $\level{.}$ is even finitely partitioning on
$B_{Tab_P}^E \cap Call(P,S)$.
This is because each (contracting) node of $\overline{\overline{CG}}$
contains only a finite number of $Tab_P$-atoms and because of the 
construction of $\level{.}$.

\item
Let $A$ be an atom such that $\tilde{A} \in Call(P,S)$,
let $H \la B_1, \ldots, B_n$ be a clause in $P$, such that $mgu(A,H)= \theta$
exists,
let $\theta_{i-1}$ be a $cas$
for $\la (B_1, \ldots, B_{i-1})\theta$:

\begin{itemize}

\item
then $\level{A} \geq \level{B_i \theta \theta_{i-1}}$.

This is because there is a directed arc from $A$ to
$B_i \theta \theta_{i-1}$ in 
$Call$-$Graph(P,S)$
and because of the construction of $\level{.}$.

\item
then
$\level{A} > \level{B_i \theta \theta_{i-1}}$ 
if $Rel(A) \simeq Rel(B_i)$ $\in NTab_P$
and $C_2(Rel(A),Rel(B_i))$ does not hold (i.e.\ $C_1(Rel(A),Rel(B_i))$
holds).

There is a directed arc in $Call$-$Graph(P,S)$ 
from $A$ to $B_i \theta \theta_{i-1}$.
Note that $A$ and $B_i \theta \theta_{i-1}$ do not belong to the same 
strongly connected component of $Call$-$Graph(P,S)$.
This is because $C_1(Rel(A),Rel(B_i))$ holds.
So, $A$ and $B_i \theta \theta_{i-1}$ 
belong to a different layer and $B_i \theta \theta_{i-1}$ is a direct 
descendant of $A$.
Hence, because of the construction of $\level{.}$, 
$\level{A} > \level{B_i \theta \theta_{i-1}}$.
\qed 
\end{itemize}
\end{itemize}
\end{proof}

\begin{example}
Recall the programs $P$ and $P^{'}$ with $Tab_{P} = Tab_{P^{'}} = \{a/0\}$ 
of Example \ref{ex:wellchosen1}.
Let $S=\{a\}$.
The LG-forests for $P \cup \{\la a\}$ and
$P^{'} \cup \{\la a\}$ are shown in Figure \ref{fig:quasi}.
\begin{figure}[hbt]
\epsfysize=6.5cm
\centerline{\epsffile{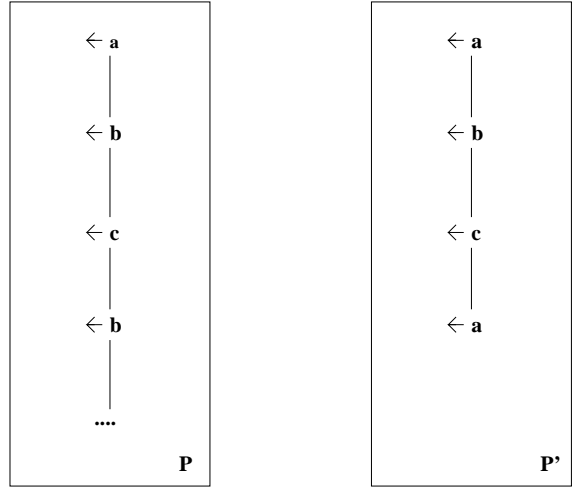}}
\caption{The LG-forests for $P \cup \{\la a\}$ and for $P^{'} \cup \{\la a\}$}
\label{fig:quasi}
\end{figure}
$P$ does not quasi-terminate w.r.t.\ $\{a/0\}$ and $S$,
whereas $P^{'}$ quasi-terminates w.r.t.\ $\{a/0\}$ and $S$.
\\
This can be proven by Theorem \ref{theoquasichar}.
Recall from Example \ref{ex:wellchosen2} that for  
both programs, the tablings are well-chosen.
Also note that, because the programs are propositional,
every level mapping is finitely partitioning on 
the whole Herbrand base.
\\
Let's first consider program $P$.
Recall that for this program and tabling $\{a/0\}$ the condition
$C_1(b,c)$ holds.
Note that there is no level mapping $\level{.}$
such that $\level{b} > \level{c}$ and $\level{c} > \level{b}$ holds.
Hence, the condition in Theorem \ref{theoquasichar} can not be satisfied
and $P$ does not quasi-terminate w.r.t.\/ $\{a/0\}$ and $S$.
\\
Consider next program $P^{'}$.
Recall that for this program and tabling $\{a/0\}$ the condition
$C_2(b,c)$ holds.
Let $\level{.}$ be the following level mapping $\level{a} = 
\level{b} = \level{c} = 0$.
With this level mapping, $P^{'}$ satisfies the condition of 
Theorem~\ref{theoquasichar} and hence, $P^{'}$ quasi-terminates 
w.r.t.\/ $\{a/0\}$ and $S$.
\end{example}

The quasi-acceptability condition is necessary only in case
the tabling is well-chosen.
We next give an example of a program $P$, a tabling $Tab_P$ which is 
not well-chosen w.r.t.\ $P$, and a set of queries $S$, such that 
$P$ quasi-terminates w.r.t.\ $Tab_P$ and $S$, but $P$ is not quasi-acceptable
w.r.t.\ $Tab_P$ and $S$.

\begin{example}
Let $P$ be the following program:
\[
\left\{
\begin{array}{lll}
p(X) & \la & q(X)
\\
q(X) & \la & r(X)
\\
r(s(X)) & \la & q(X)
\\
r(X) & \la & p(X)
\end{array}
\right.
\]
with tabling $Tab_P = \{p/1\}$.
Notice that $Tab_P$ is not well-chosen w.r.t.\ $P$.
Let $S = \{p(0)\}$.
$P$ quasi-terminates w.r.t.\ $Tab_P$ and $S$.
We show that $P$ is not quasi-acceptable w.r.t.\ $Tab_P$ and $S$.
Suppose that there exists a level mapping $\level{.}$ such that 
$P$ is quasi-acceptable w.r.t.\ $Tab_P$, $S$ and this level mapping
$\level{.}$ (we prove a contradiction).
Then, for this level mapping, the following inequalities must hold:
$\level{p(0)} \geq \level{q(0)}$, $\level{q(0)} > \level{r(0)}$ 
(since $C_3(q,r)$ holds, and so $C_2(q,r)$ does not hold), and
$\level{r(0)} \geq \level{p(0)}$.
Hence, $\level{p(0)} > \level{p(0)}$ must hold, but this gives a
contradiction.
\end{example}

\subsection{LG-Termination} \label{ssectlgchar}

In analogy to quasi-termination, we now present
a necessary and sufficient condition for LG-termination in case the
tabling is well-chosen.  In case the tabling is not well-chosen, the
condition is still sufficient.

Note that Theorem~\ref{theolg} already provides us with a 
characterisation of LG-termination of a program 
in terms of quasi-termination.
That is, to prove the LG-termination of $P$ w.r.t.\ $Tab_P$ and $S$, 
it suffices
to prove the quasi-termination of $P^a$, the a-transformation
of the program $P$, w.r.t.\ $Tab_{P^a}$ and $S$.
To prove quasi-termination, we can use the results of
Subsection~\ref{ssectquasichar}.
Namely, it is sufficient (and also necessary in case the tabling
is well-chosen\footnote{
Note that if $Tab_P$ is well-chosen w.r.t.\ $P$, then also
$Tab_{P^a}$ is well-chosen w.r.t.\ $P^a$.}) to prove the 
quasi-acceptability of $P^a$ w.r.t.\ $Tab_{P^a}$ and $S$.  
However, the condition of quasi-acceptability on $P^a$ can 
be weakened; i.e.\ 
some of the decreases ``$\level{A} \geq \level{B_i\theta\theta_{i-1}}$''
need not be checked because
they can always be fulfilled.
In particular, we only have to require the non-strict decrease
for recursive, tabled body atoms $B_i$ (to obtain an LG-forest with
only finitely many LG-trees) or for body atoms $B_i$ of the form
$p^a(t_1,\ldots,t_n)$ (to obtain LG-trees which are finitely branching);
the conditions on non-tabled predicates remain the same.
The following notion of LG-acceptability gives this
optimised condition for
LG-termination of a program.

\begin{definition}[LG-acceptability] \label{deflgacce}
Let $P$ be a program, $Tab_P \subseteq Pred_P$
and $S \subseteq B_P^E$.
$P$ is {\em LG-acceptable} w.r.t.\ $Tab_P$ and $S$ iff
\\
there is a level mapping $\level{.}$ on $B_{P^a}^E$  such that
for all $A$ such that $\tilde{A} \in S$, $\level{.}$
is finitely partitioning on 
$Call(P^{a},\{A\}) 
\cap B^E_{TR_P \cup TR_P^{a}}$,
and such that

\begin{itemize}

\item
for every atom $A$ such that $\tilde{A} \in Call(P^a,S)$,

\item

for every clause $H \la B_1, \ldots, B_n$ in $P^{a}$, 
such that $mgu(A,H)= \theta$
exists,

\item
for every $B_i$ such that $Rel(B_i) \simeq Rel(H)$ or
$Rel(B_i) \in TR^{a}_P$,

\item
for every $cas$ $\theta_{i-1}$ 
in $P^{a}$
for $\la (B_1, \ldots, B_{i-1})\theta$:

\[
\begin{array}{ll}
\level{A} \geq  \level{B_i \theta \theta_{i-1}} &
\\
\mbox{and} &
\\
\level{A} >  \level{B_i \theta \theta_{i-1}} & \mbox{if}~Rel(A) \simeq Rel(B_i) 
\in NTab_P~\mbox{and}
\\
 & C_2(Rel(A),Rel(B_i))~\mbox{does not hold} .
\end{array}
\]
\end{itemize}
\end{definition}

\begin{theorem}[(necessary and) sufficient condition for
LG-termination] \label{theolgchar}
Let $P$ be a program, $Tab_P \subseteq Pred_P$ and $S \subseteq B^E_P$.
\\
If $P$ is LG-acceptable w.r.t.\ $Tab_P$ and $S$, then $P$
LG-terminates w.r.t.\ $Tab_P$ and $S$.
\\
If the tabling $Tab_P$ is well-chosen w.r.t.\ $P$, then also the converse
holds, i.e.\ $P$ is LG-acceptable w.r.t.\ $Tab_P$ and $S$
iff $P$ LG-terminates w.r.t.\ $Tab_P$ and $S$.
\end{theorem}

\begin{proof}
$\Rightarrow :$ Suppose that $P$ is LG-acceptable w.r.t.\ $Tab_P$ and $S$.
We prove that $P$
LG-terminates w.r.t.\ $Tab_P$ and $S$.
\\
Let $A$ be an atom such that $\tilde{A} \in S$. Let $\F$ be the LG-forest
w.r.t.\ $Tab_P$ of $P \cup \{\la A\}$. We prove that $\F$ consists of a 
finite number of finite LG-trees.

\begin{itemize}
\item
The LG-trees in $\F$ are finitely branching.

Suppose this is not the case, i.e.\ there is an LG-tree in $\F$ which is 
infinitely branching.
Then, there is an LG-tree $T$ in $\F$ which is infinitely branching 
in a non-root
node, which is a query with leftmost atom 
$p(t_1,\ldots,t_n)$, with $p \in TR_P$, which is directly descending
from an atom $q(s_1,\ldots,s_m)$, with $p \simeq q$, via a 
recursive clause $C =q(u_1,\ldots,u_m) \la \ldots, p(v_1,\ldots,v_n), \ldots$.
Now, consider the LG-forest $\F^a$ of $P^a \cup \{\la A\}$.
Let $T^a$ be the LG-tree in $\F^a$ corresponding to $T$.
Note that the clause $C^a$ instead of $C$ is used in $T^a$.
Because of this, the atom on the right of 
$p(t_1,\ldots,t_n)$ in 
the infinitely branching node 
is $p^a(t_1,\ldots,t_n)$. Thus, 
$\F^a$ consists of a infinite number of LG-trees (there are an infinite number
of LG-trees with predicate $p^a$ in the root).
But, all these $p^a$-atoms directly descend from the node 
$q(s_1,\ldots,s_m)$ via the clause $C^a$ in $P^a$ and hence,
because of the LG-acceptability condition,
their value under the level mapping $\level{.}$
is smaller or equal to 
$\level{q(s_1,\ldots,s_m)}$.
Because $\level{.}$ is finitely partitioning on 
$Call(P^a,\{A\}) \cap B^E_{TR_P^a}$, this gives a contradiction.

\item
$\F$ consists of a finite number of LG-trees, i.e.\ 
$\sharp( Call(P,\{A\}) \cap B^E_{Tab_P} ) <
\infty$.

Suppose this is not the case. 
A first possible reason for an infinite number of LG-trees in $\F$ is
an infinitely branching LG-tree in $\F$. But we already proved that this 
does not occur.
The other possibility is
that there exists an 
infinite LD-derivation of $\la A$
in $P$ which contains an infinite directed subsequence,
such that this infinite directed subsequence has a tail
$G_n, G_{n+1}, \ldots$ with $G_i = \la A_i, \A_i$, $i\geq n$, such that
$\{\tilde{A}_i~|~i \geq n\} \subseteq B^E_{Tab_P}$ 
is an infinite set and $Rel(A_i) \simeq Rel(A_{i+1})$ for
all $i\geq n$. So, $\{\tilde{A}_i~|~i \geq n\} \subseteq B^E_{TR_P}$. 
Since $\tilde{A}_i \in Call(P,\{A\}) \cap B^E_{TR_P}$ $\subseteq$ 
$Call(P^a,\{A\}) \cap B^E_{TR_P \cup TR_P^a}$, 
and since $\level{.}$ is finitely partitioning
on this set
and $\level{A_i} \geq \level{A_{i+1}}$ for all $i \geq n$
(by the LG-acceptability condition), this
gives a contradiction.

\item The LG-trees in $\F$ have finite branches.

The same argumentation as in the proof of Theorem \ref{theoquasichar} can
be applied here.
\end{itemize}

\noindent
$\Leftarrow :$
Suppose that 
the tabling $Tab_P$ is well-chosen w.r.t.\ $P$ and suppose that
$P$ LG-terminates w.r.t.\ $Tab_P$ and $S$.
We prove that there exists a level mapping $\level{.}$ such that
$P$ is LG-acceptable w.r.t.\ $Tab_P$, $S$ and this level mapping
$\level{.}$.
\\
Since $P$ LG-terminates w.r.t.\ $Tab_P$ and $S$,
we know by Theorem~\ref{theolg} that $P^a$ quasi-terminates 
w.r.t.\ $Tab_{P^a}$ and $S$.
Note that, since $Tab_P$ is well-chosen w.r.t.\ $P$,
$Tab_{P^a}$ is well-chosen w.r.t.\ $P^a$.
By Theorem~\ref{theoquasichar}, there exists a level mapping
$\level{.}$ such that $P^a$ is quasi-acceptable w.r.t.\ $Tab_{P^a}$,
$S$ and this level mapping $\level{.}$.
It is straightforward to verify that $P$ is LG-acceptable w.r.t.\ $Tab_P$,
$S$ and this level mapping $\level{.}$.
(Note that, as we already discussed in the beginning of this subsection,
the level mapping obtained in this way satisfies more
conditions than required by the notion of LG-acceptability.)
\qed
\end{proof}

\begin{example} \label{exagrammarlg}
Recall the part $R$ of the grammar program (Section~\ref{ssectmotexa}) which
recognizes the language $a^nb$:
\[
\begin{array}{ll}
R: &
\left\{
\begin{array}{lll}
  s(Si,So) & \la & a(Si,S), S = [b|So] \\
  a(Si,So) & \la & a(Si,S), a(S,So)    \\
  a(Si,So) & \la & Si = [a|So]
\end{array}
\right.
\end{array}
\]
with $Tab_R = \{a/2\}$.  We show that $R$ LG-terminates w.r.t.\/
$\{a/2\}$ and $S = \{s(si,So)~|~si$ is a ground list consisting of
constants $a,b$ and $So$ is a variable$\}$.  Consider the
a-transformation of $R$:
\[
\begin{array}{ll}
R^a: &
\left\{
\begin{array}{lll}
  s(Si,So) & \la & a(Si,S), S = [b|So] \\
  a(Si,So) & \la & a(Si,S), a^a(Si,S), a(S,So), a^a(S,So) \\
  a(Si,So) & \la & Si = [a|So] \\
  a^a(Si,So) & \la &
\end{array}
\right.
\end{array}
\]
with $Tab_{R^a} = \{a/2,a^a/2\}$.
When applying Theorem~\ref{theolgchar}, we only have to consider
the second clause of $R^a$.
Note that, for all $a(t1,t2) \in Call(R^a, \{s(si,So)\})$,
$t1 $ is a sublist of $si$ and $t2$ is a variable.
Also, for all $a^a(v1,v2) \in Call(R^a, \{s(si,So)\})$,
$v1$ is a sublist of $si$ and $v2$ is a (strict) sublist of $v1$.
Let $\level{.}$ be the following level mapping:
\[
\begin{array}{ll}
\level{a(t1,t2)} & = 2 \norm{t1}_l
\\
\level{a^a(v1,v2)} & = \norm{v1}_l + \norm{v2}_l
\end{array}
\]
where $\norm{.}_l$ is the list-length norm\footnote{The list-length
norm is defined as follows: 
\[ 
\left\{
\begin{array}{lll}
\norm{[h|t]}_l & = 1 + \norm{t}_l &
\\
\norm{u}_l & = 0 & \quad \mbox{if}~u \not= [h|t].
\end{array}
\right.
\]
}.
The level mapping $\level{.}$ is finitely partitioning on the
whole set $Call(R^a,S) \cap B^E_{\{a/2,a^a/2\}}$.
It can be easily verified that
$R$ and $S$, together with $\level{.}$, satisfy the conditions of 
Theorem \ref{theolgchar}.
Hence, $R$ LG-terminates w.r.t.\ $\{a/2\}$ and $S$.
\end{example}

\section{Modular Termination Proofs for Tabled Logic Programs} 
\label{sectmod}

In the context of 
programming in the large,
it is important to be able to obtain modular termination proofs, i.e.\
proofs built by combining termination proofs of separate components of
the program.  Starting from the quasi- and LG-acceptability
conditions, we present modular proofs for quasi-termination in
Subsections~\ref{ssectquasimod} and \ref{ssectconstrlm}, and for
LG-termination in Subsection~\ref{ssectlgmod}.  We consider the union
$P\cup R$ of two programs $P$ and $R$, where $P$
extends\footnote{Recall that a program $P$ {\em extends} a program $R$
iff no predicate defined in $P$ occurs in $R$.}  $R$, and we prove the
quasi/LG-termination of $P \cup R$ by imposing conditions on the two
components $P$ and $R$.

In order to fix a notation, for 
$Pred_{P\cup R} = Tab_{P\cup R} \sqcup NTab_{P\cup R}$, let
\[
\begin{array}{lll}
Tab_P = Tab_{P\cup R} \cap Pred_P & , &
NTab_P = NTab_{P\cup R} \cap Pred_P
\\ 
Tab_R = Tab_{P\cup R} \cap Pred_R & , &
NTab_R = NTab_{P\cup R} \cap Pred_R .
\end{array}
\]
So the tabling of the union $P \cup R$ determines the tabling of the
components $P$ and $R$. 
Note that $Tab_P$ also contains predicates
which are tabled in $P\cup R$ but defined in $R$. 

In the following we give modular termination proofs for the union
$P\cup R$ of two programs $P$ and $R$ where:
\begin{enumerate}
\item \label{itemmod1}
  $P$ extends $R$.
\item \label{itemmod2}
  $P$ extends $R$ and no defined predicate in $P$ is tabled 
  ($Def_P \subseteq NTab_P$).
\item \label{itemmod3}
  $P$ extends $R$ and all defined predicates in $P$ are tabled
  ($Def_P \subseteq Tab_P$).
\item \label{itemmod4}
  $P$ extends $R$ and $R$ extends $P$.
\end{enumerate}
Note that points \ref{itemmod2}, \ref{itemmod3} and \ref{itemmod4} are
special cases of the first one.
The reason for treating them separately is because they occur quite
often in practice and, more importantly, because
in these special cases, simpler modular termination conditions can be given.

\subsection{Modular Conditions for Quasi-Termination}
\label{ssectquasimod}

Throughout this subsection, we will consider the following example.

\begin{example} \label{examodular}
Consider the following union of programs $U = T \cup P \cup R \cup P^{'}$
with $Tab_U = \{path/4\}$.
Let $S = \{reachable(rome,X)\}$, then $U$ will compute 
the cities belonging to the same region $r$ as $rome$ and which are
reachable from $rome$ making use of 
the list of connections of the region $r$.
The program $P^{'}$ contains facts
giving the region to which each city belongs and 
the 
list of connections in each region (a connection between city $c_1$ and
city $c_2$ is given by the term $e(c_1,c_2)$). 
\[
\begin{array}{ll}
T: & \left\{
\begin{array}{lll}
reachable(X,Y) & \la & inregion(X,R),\ connections(R,Ed), 
\\
  & &                   path(X,Ed,Y,L)
\end{array}
\right.
\\
~
\\
P: & \left\{
\begin{array}{lll}
path(X,Ed,Y,[Y]) & \la & edge(X,Ed,Y)
\\
path(X,Ed,Z,[Y|L]) & \la & edge(X,Ed,Y),\ path(Y,Ed,Z,L)
\end{array}
\right.
\\
~
\\
R: & \left\{
\begin{array}{lll}
edge(X,[e(X,Y)|L],Y) & \la & 
\\
edge(X,[e(X_1,X_2)|L],Y) & \la & edge(X,L,Y)
\end{array}
\right.
\\
~
\\
P^{'}: & \left\{
\begin{array}{lll}
inregion(city,region) . & &
\\
\ldots
\\
connections(region,list\_of\_connections). & &
\\
\ldots
\end{array}
\right.
\end{array}
\]
We will prove that $U$ quasi-terminates w.r.t.\ 
$Tab_U$ and $S$.
We will do this in a modular way:
\begin{itemize}
\item
In Example \ref{exaqmod1}, after Proposition \ref{propqmod1}, we prove that
$U = T \cup (P \cup R \cup P^{'})$ quasi-terminates, 
given that $(P \cup R \cup P^{'})$
quasi-terminates.

\item
In Example \ref{exaqmod3}, after Proposition \ref{propqmod3}, we prove that
$P \cup R$ quasi-terminates (recall that $P \cup R$ is the program of
Example~\ref{exapath} in Subsection~\ref{ssectslg}).

\item
In Example \ref{exaqmod4}, after Proposition \ref{propqmod4}, we prove that
$(P \cup R) \cup P^{'}$ quasi-terminates, given that $(P\cup R)$
and $P^{'}$  quasi-terminate.
\end{itemize}
 
\end{example}

\begin{proposition} \label{propqmod1}
Suppose $P$ and $R$ are two programs, such that $P$ extends $R$.
Let $S \subseteq B_{P \cup R}^E$.
If

\begin{itemize}
\item $R$ quasi-terminates w.r.t.\ $Tab_R$ and $Call(P \cup R,S)$,

\item there is a level mapping $\level{.}$ on 
$B_{P}^E$ such that 
for all $A$ such that $\tilde{A} \in S$, $\level{.}$ 
is finitely partitioning on 
$Call(P\cup R,\{A\}) \cap B_{Tab_{P}}^E$,
and such that

\begin{itemize}
\item
for every atom $A$ such that $\tilde A \in Call(P\cup R,S)$,

\item
for every clause $H \la B_1, \ldots,B_n$ in $P$ such that
$mgu(A,H) = \theta$ exists,

\item
for every $1 \leq i \leq n$,

\item
for every $cas$ $\theta_{i-1}$ in $P \cup R$ 
for $\la (B_1,\ldots,B_{i-1})\theta$:

\end{itemize}

\[
\begin{array}{ll}
\level{A} \geq \level{B_i \theta \theta_{i-1}} &
\\
\mbox{and} &
\\
\level{A} >  \level{B_i \theta \theta_{i-1}} & \mbox{if}~Rel(A) \simeq Rel(B_i) 
\in NTab_P~ \mbox{and}
\\
 & C_2(Rel(A),Rel(B_i))~\mbox{does not hold} .
\end{array}
\]
\end{itemize}
then, $P \cup R$ quasi-terminates w.r.t.\ $Tab_{P\cup R}$ and $S$.
\end{proposition}

\begin{proof}
Let $A$ be an atom such that $\tilde{A} \in S$.
Let $\F$ be the LG-forest w.r.t.\ $Tab_{P\cup R}$ of
$P\cup R \cup \{\la A\}$.
We prove that $\F$ consists of a finite number of LG-trees without
infinite branches.
\\
If $A$ is defined in $R$, this follows directly from the fact that
$P$ extends $R$ and that
$R$ quasi-terminates w.r.t.\ $Tab_R$ and $Call(P\cup R,S)$.
\\
So, suppose $A$ is defined in $P$.
Because of the second condition in the proposition statement,
every call directly descending from $A$, say $B$, is such that
$\level{B} \leq \level{A}$. 
This holds recursively for atoms descending from $A$ using
clauses of $P$.
Because $\level{.}$ is finitely partitioning on $B_{Tab_{P}}^E \cap
Call(P\cup R,\{A\})$, the set of tabled atoms, descending from $A$, using
clauses of $P$, is finite.
For atoms $C$, defined in $R$ and descending from $A$
using clauses of $P$, we 
know that $\sharp (Call(P\cup R,\{C\}) \cap B^E_{Tab_R}) < \infty$.
So, $\sharp (Call(P\cup R,\{A\}) \cap B_{Tab_{P\cup R}}^E) < \infty$.
\\
We now prove that there is no tree in $\F$ with an infinite branch.
Suppose this is not the case, and there is a tree in $\F$ with an infinite 
branch. 
Because, $R$ quasi-terminates w.r.t.\ $Tab_R$ and $Call(P\cup R,S)$,
and because $P$ extends $R$,
this infinite branch contains an infinite directed subsequence
$G_0, G_1, \ldots$, with leftmost atoms $A_0, A_1, \ldots$, 
belonging to $B_{NTab_P \cap Def_P}^E$.
This infinite directed subsequence has a tail, 
such that for all $i$ such that
$G_i$ belongs to this tail,
$Rel(A_i) \simeq Rel(A_{i+1})$ and
$C_2(Rel(A_i),Rel(A_{i+1}))$ does not hold.
But because of the condition in the proposition statement,
$\level{A_i} > \level{A_{i+1}}$ and this gives a contradiction. 
\qed \end{proof}

\begin{example}[Example \ref{examodular} continued] \label{exaqmod1}
We illustrate the above proposition by proving that
$U = T \cup (P \cup R \cup P^{'})$ quasi-terminates w.r.t.\ 
$Tab_U = \{\mathit{path}/4\}$ and $S =
\{\mathit{reachable(rome,X)}\}$, given that
$P \cup R \cup P^{'}$ quasi-terminates w.r.t.\ $\{\mathit{path}/4\}$ and 
$Call(U,S)$.
The quasi-termination of $P \cup R \cup P^{'}$ will be shown in the
following examples of this subsection.
\\
The trivial level mapping (mapping every atom to $0$) satisfies the
condition of the proposition; there is no recursive call to a non-tabled
predicate in $T$ and the set of called $\mathit{path}$-atoms is finite
(since the database $P^{'}$ is finite). 
\end{example}

The case of two programs $P$ and $R$, such that $P$ extends $R$ and
such that no defined predicate in $P$ is tabled (mentioned as
point~\ref{itemmod2} in the introduction of Section~\ref{sectmod}),
does not give rise to a simpler modular termination condition than the
condition in Proposition~\ref{propqmod1}. We want to note already here
that regarding LG-termination, this special case
(point~\ref{itemmod2}) will give rise to a simpler modular termination
condition than in the general case.

The next proposition considers the special case of two programs $P$ and $R$,
such that $P$ extends $R$ and such that 
all the defined predicates in $P$ are tabled.

\begin{proposition} \label{propqmod3}
Suppose $P$ and $R$ are two programs, such that $P$ extends $R$,
and such that $Def_P \subseteq Tab_P$.
Let $S \subseteq B_{P \cup R}^E$.
If

\begin{itemize}

\item
$R$ quasi-terminates w.r.t.\ $Tab_R$ and $Call(P \cup R,S)$,

\item
there is a level mapping $\level{.}$ on $B_{P}^E$ such that
for all $A$ such that $\tilde{A} \in S$, $\level{.}$
is finitely partitioning 
on $Call(P\cup R,\{A\}) \cap B_{Tab_{P}}^E$, 
and such that

\begin{itemize}

\item
for every atom $A$ such that $\tilde A \in Call(P\cup R,S)$,

\item
for every clause $H \la B_1, \ldots,B_n$ in $P$ such that
$mgu(A,H) = \theta$ exists,

\item
for every $1 \leq i \leq n$,

\item
for every $cas$ $\theta_{i-1}$ in $P \cup R$ 
for $\la (B_1,\ldots,B_{i-1})\theta$:

\[
\level{A} \geq \level{B \theta \theta_{i-1}}
\]

\end{itemize}
\end{itemize}
then, $P \cup R$ quasi-terminates w.r.t.\ $Tab_{P\cup R}$ and $S$.
\end{proposition}

\begin{proof}
This is a direct corollary of Proposition \ref{propqmod1} (every
recursive predicate in $P$ is defined in $P$ and hence tabled).
\qed
\end{proof}

\begin{example}[Example \ref{examodular} continued] \label{exaqmod3}
We illustrate the above proposition by proving that 
$P \cup R$ quasi-terminates w.r.t.\ $\{\mathit{path}/4\}$ and
$Call(U,S) \cap B^E_{P\cup R}$.
\begin{itemize}
\item First we prove that $R$ quasi-terminates w.r.t.\ $\emptyset$ and
      $Call(U,S) \cap B^E_{R}$ (or, since there are no tabled atoms in
      $R$, that $R$ LD-terminates w.r.t.\ $Call(U,S) \cap B^E_{R}$).
      We use Theorem~\ref{theoquasichar} and show that $R$ is
      quasi-acceptable w.r.t.\ $\emptyset$ and $Call(U,S) \cap B^E_{R}$.
      Consider the following level mapping:
      \[
        \level{edge(t_1,t_2,t_3)} = \norm{t_2}_l
      \]
      It can be easily seen that we have a strict decrease between the
      head and the body atom of the recursive clause for $edge$ in $R$.
      Hence, the quasi-acceptability condition is satisfied.
\item The trivial level mapping on $B^E_P$ satisfies the second
      condition in the proposition statement.  Indeed, $\mathit{path}$
      is tabled so a strict decrease is never required, and the set of
      called $\mathit{path}$-atoms is finite since the database of
      facts comprising $P^{'}$ is finite.
\end{itemize}
\end{example}

Finally, we consider the case of two programs $P_1$ and $P_2$ extending
each other.

\begin{proposition} \label{propqmod4} 
Let $P_1, P_2$ be two programs such that $P_1$ extends $P_2$ and 
$P_2$ extends $P_1$.
Let $S \subseteq B^E_{P_1 \cup P_2}$.
If

\begin{itemize}

\item 
$P_1$ quasi-terminates w.r.t.\ $Tab_{P_1}$ and $S \cap B_{P_1}^E$,

\item 
$P_2$ quasi-terminates w.r.t.\ $Tab_{P_2}$ and $S \cap B_{P_2}^E$,

\end{itemize}
then 
$P_1 \cup P_2$ quasi-terminates w.r.t.\ $Tab_{P_1\cup P_2}$ and $S$. 

\end{proposition}

\begin{proof}
Because $P_1$ extends $P_2$ and 
$P_2$ extends $P_1$, 
$Call(P_1\cup P_2, S) \cap B^E_{P_i} = Call(P_i, S\cap B^E_{P_i})$ 
for $i =1,2$.   
The proposition follows then by definition of quasi-termination.
\qed \end{proof}

\begin{example}[Example \ref{examodular} continued] \label{exaqmod4}
We prove that $(P \cup R) \cup P^{'}$ quasi-terminates w.r.t.\
$\{\mathit{path}/4\}$ and $Call(U,S) \cap B^E_{P\cup R \cup P^{'}}$,
given that $P \cup R$ quasi-terminates w.r.t.\ $\{\mathit{path}/4\}$
and $Call(U,S) \cap B^E_{P\cup R}$ (which was shown in
Example~\ref{exaqmod3}) and that $P^{'}$ quasi-terminates w.r.t.\
$\{\mathit{path}/4\}$ and $Call(U,S) \cap B^E_{P^{'}}$ (which is
obvious since it consists of a finite set of facts).  We can apply
Proposition~\ref{propqmod4}, since $P \cup R$ extends $P^{'}$ and vice
versa, $P^{'}$ extends $P \cup R$.
\end{example}

The above modular conditions for the quasi-termination of $P\cup R$ 
are proven to be sufficient, but in many cases they are also necessary.
In particular, the modular conditions of Proposition~\ref{propqmod1} are
also necessary for the quasi-termination of $P\cup R$ in case the tabling
$Tab_P$ is well-chosen w.r.t.\ $P$.
Because in all cases where $Def_P \subseteq Tab_P$, $Tab_P$
is well-chosen w.r.t.\ $P$, it follows that the modular conditions of
Proposition~\ref{propqmod3} are necessary in general.
Finally, it can be easily seen that the modular conditions of 
Proposition~\ref{propqmod4} are also necessary in general.

Note that all the above modular termination conditions prove the
quasi-termination of $P \cup R$ without constructing a
level mapping $\level{.}$ such that $P\cup R$ is quasi-acceptable
w.r.t.\ this level mapping.
In Subsection \ref{ssectconstrlm}, modular termination conditions
for quasi-termination are given which construct (from simpler
level mappings) a level mapping
such that $P\cup R$ is quasi-acceptable w.r.t.\ this level mapping.
This construction will be illustrated in Example~\ref{exaconstrlm}
on the program $P\cup R$ of
Example~\ref{examodular} (see also Example~\ref{exaqmod3}).
 
\subsection{Modular Conditions for LG-Termination}
\label{ssectlgmod}

Similarly to the case of quasi-termination, we want to have modular
termination proofs for the LG-termination of the union $P\cup R$ of
two programs $P$ and $R$, where $P$ extends $R$.  Note that, because
of Theorem~\ref{theolg} and because $(P\cup R)^{a} =$ $P^{a} \cup
R^{a}$ (if $P$ extends $R$), we can use the modular proofs for
quasi-termination of Subsection~\ref{ssectquasimod}.  However, as we
already noted in Subsection~\ref{ssectlgchar}, we can give simpler
conditions which require less checks of decreases between the levels
of successive calls.  These conditions are given below.

\begin{proposition} \label{propmodlg}
Let $P$ and $R$ be two programs, such that $P$ extends $R$.
Let $S \subseteq B_{P \cup R}^E$.
If

\begin{itemize}
\item $R$ LG-terminates w.r.t.\ $Tab_R$ and $Call(P\cup R,S)$, and

\item there is a level mapping $\level{.}$ on $B^E_{P^a}$ such that
      for all $A$ such that $\tilde{A} \in S$, $\level{.}$
      is finitely partitioning on  
      $Call(P^a \cup R, \{A\}) \cap B_{TR_P \cup TR^a_P}^E$,
      and such that
      \begin{itemize}
      \item for every atom $A$ such that
            $\tilde A \in Call(P^a \cup R, S)$,
      \item for every clause $H \la B_1, \ldots, B_n$ in $P^a$ 
            such that $mgu(A, H) = \theta$ exists,
      \item for every $B_i$ such that $Rel(B_i) \simeq Rel(H)$ or
            $Rel(B_i) \in TR^a_P$,
      \item for every $cas$ $\theta_{i-1}$ in $P^a \cup R$ for 
            $\la (B_1,\ldots,B_{i-1})\theta$:
                \[
                \begin{array}{ll}
                \level{A} \geq \level{B_i\theta \theta_{i-1}} &
                \\
                \mbox{and} &
                \\
                \level{A} > \level{B_i\theta \theta_{i-1}} & 
                \mbox{if}~Rel(A) \simeq Rel(B_i) \in NTab_P~\mbox{and}
                \\
                 & C_2(Rel(A),Rel(B_i))~\mbox{does not hold} .
                \end{array}
                \]
      \end{itemize} 
\end{itemize}
then $P\cup R$ LG-terminates w.r.t.\ $Tab_{P\cup R}$ and $S$.
\end{proposition}

\begin{proof}
The proof is a simple adaptation of the proof of the if-direction
of Theorem~\ref{theolgchar}; the adaptation is similar to the 
adaptation needed to transform the proof of the if-direction
of Theorem~\ref{theoquasichar} into a proof of Proposition~\ref{propqmod1}.
\qed
\end{proof}

We next consider three special cases of Proposition~\ref{propmodlg}.
In the following proposition we consider the case in which no defined
predicate in $P$ is tabled.

\begin{proposition} \label{proplgmod2}
Let $P$ and $R$ be two programs, such that $P$ extends $R$
and such that $\mathit{Def}_P \subseteq \mathit{NTab}_P$.
Let $S \subseteq B_{P \cup R}^E$.
If
\begin{itemize}
\item $R$ LG-terminates w.r.t.\ $\mathit{Tab}_R$ and $Call(P\cup R,S)$,
\item there is a level mapping $\level{.}$ on $B_P^E$ such that
      \begin{itemize}
      \item for every atom $A$ such that $\tilde A \in Call(P\cup R,S)$,
      \item for every clause $H\la B_1,\ldots,B_n$ in $P$ such that
            $mgu(A,H)=\theta$ exists,
      \item for every $B_i$ such that $Rel(B_i) \simeq Rel(A)$,
      \item for every $cas$ $\theta_{i-1}$ in $P\cup R$ for 
            $\la (B_1,\ldots,B_{i-1}) \theta$:
                $$
                \level{A} > \level{B_i\theta \theta_{i-1}}
                $$
      \end{itemize}
\end{itemize}
then $P\cup R$ LG-terminates w.r.t.\ $\mathit{Tab}_{P\cup R}$ and $S$.
\end{proposition}

\begin{proof}
Because no defined predicate in $P$ is tabled, $P^a = P$.
Also, for all $p, q \in \mathit{NTab}_P \cap \mathit{Def}_P$ with $p
\simeq q$, $C_1(p,q)$ holds.  
The proposition follows then from Proposition \ref{propmodlg}.
\qed 
\end{proof}

\begin{example}
Recall program $R$ of Example \ref{exagrammarlg}.
Let $P$ be the following program which parses the language $a^nb$
(see also Subsection~\ref{ssectmotexa}):
\[
\begin{array}{ll}
P: &
\left\{
\begin{array}{lll}
\mathit{s(Si,So,PT)} & \la & \mathit{a(Si,S), S = [b|So],
                                     PT = spt(PTa,b), a(Si,S,PTa)} \\
\mathit{a(Si,So,PT)} & \la & \mathit{a(Si,S), a(S,So),
                                     PT = apt(PT1,PT2), a(Si,S,PT1),} \\
                     &     & \mathit{a(S,So,PT2)} \\
\mathit{a(Si,So,PT)} & \la & \mathit{Si = [a|So], P T= a}
\end{array}
\right.
\end{array}
\]
As already noted, $P$ extends $R$.  Let $a/2$ be the only tabled
predicate in $P\cup R$; see Subsection~\ref{ssectmotexa} for why this
tabling is sufficient.
Let $S = \{\mathit{s(si,So,PT)}~|~si$ is a ground list
consisting of constants $a,b$, and $So, PT$ are 
distinct variables$\}$.
We show, using Proposition~\ref{proplgmod2}, that $P\cup R$
LG-terminates w.r.t.\/ $\{a/2\}$ and $S$.
\begin{itemize}
\item $R$ LG-terminates w.r.t.\ $\{a/2\}$ and $Call(P\cup R,S)$.

      Note that, if $a(t1,t2) \in Call(P\cup R,s(si,So,PT))$, then either $t1$
      is a sublist of $si$ and $t2$ is a variable, or $t1$ and $t2$
      are both sublists of $si$.
      In Example~\ref{exagrammarlg}, we proved that $R$ LG-terminates
      w.r.t.\/ this first kind of queries. 
      To prove that $R$ LG-terminates w.r.t.\ the second kind of 
      queries, we can again apply Theorem \ref{theolgchar}. 
      Since the proof is similar to the one given in 
      Example~\ref{exagrammarlg}, we omit it here.
      
\item Note first that, if $a(t1,t2,P) \in  
      Call(P\cup R,\{s(si,So,PT)\})$, 
      then
      $t2$ is a (strict) sublist of $t1$, $t1$ is a sublist of $si$
      and $P$ is a variable.
      Let $\level{.}$ be the following level mapping on 
      $Call(P\cup R,S) \cap B^E_{\{a/3\}}$:
      $\level{a(t1,t2,P)} = \norm{t1}_l - \norm{t2}_l$.
      Because of the remark above, $\level{.}$ is well-defined. 
      Note that we only have to consider the recursive clause
      for $a/3$ in the analysis.
      \begin{itemize}
      \item First consider the fourth body atom in the recursive
            clause for $a/3$.  If this clause is called with
            $a(ti,to,PT)$, with $to$ a (strict) sublist of $ti$, then
            the fourth body atom is called as $a(ti,t,PT1)$ where $to$
            is a (strict) sublist of $t$ and $t$ is a (strict) sublist
            of $ti$.  Hence,
            $$
                \level{a(ti,to,PT)} = \norm{ti}_l - \norm{to}_l >
                \norm{ti}_l - \norm{t}_l = \level{a(ti,t,PT1)}.
            $$
      \item Now consider the last body atom.
            If the recursive clause is called with $a(ti,to,PT)$, with
            $to$ a (strict) sublist of $ti$, then the last body atom
            is called as $a(t,to,PT2)$ where $to$ is a (strict) sublist
            of $t$ and $t$ is a (strict) sublist of $ti$.  Hence,
            $$
                \level{a(ti,to,PT)} = \norm{ti}_l - \norm{to}_l >
                \norm{t}_l - \norm{to}_l = \level{a(t,to,PT2)}.
            $$
\end{itemize}
We conclude that $P\cup R$ and $S$ satisfy the condition of 
Proposition~\ref{proplgmod2}, so 
$P\cup R$ LG-terminates w.r.t.\ $\{a/2\}$ and $S$.
\end{itemize}
\end{example}

In the next proposition, a modular termination proof for the
LG-termination of the union $P\cup R$ is given, where $P$ extends $R$
and all defined predicates in $P$ are tabled.

\begin{proposition} \label{proplgmod3}
Let $P$ and $R$ be two programs, such that $P$ extends $R$
and such that $\mathit{Def}_P \subseteq \mathit{Tab}_P$.
Let $S \subseteq B_{P \cup R}^E$.  If
\begin{itemize}
\item $R$ LG-terminates w.r.t.\ $Tab_R$ and $Call(P\cup R,S)$, and
\item there is a level mapping $\level{.}$ on $B_{P^a}^E$ such that
      for all $A$ such that $\tilde{A} \in S$, $\level{.}$ is finitely
      partitioning on $Call(P^a \cup R,\{A\}) \cap B^E_{TR_P \cup TR_P^a}$,
      and such that
      \begin{itemize}
      \item for every atom $A$ such that
            $\tilde A \in Call(P^a \cup R, S)$,
      \item for every clause $H \la B_1, \ldots, B_n$ in $P^a$ 
            such that $mgu(A, H) = \theta$ exists,
      \item for every $B_i$ such that $Rel(B_i) \simeq Rel(H)$ or
            $Rel(B_i) \in TR^a_P$,
      \item for every $cas$ $\theta_{i-1}$ in $P^a \cup R$ for 
            $\la (B_1,\ldots,B_{i-1})\theta$:
            $$
                \level{A} \geq \level{B_i\theta \theta_{i-1}} 
            $$
      \end{itemize} 
\end{itemize}
then $P\cup R$ LG-terminates w.r.t.\ $\mathit{Tab}_{P\cup R}$ and $S$.
\end{proposition}

\begin{proof}
This is a direct corollary of Proposition~\ref{propmodlg}
(every recursive predicate in $P$ is defined in $P$ and hence tabled).
\qed
\end{proof}

Finally, we consider the case of two programs $P_1$ and $P_2$ 
extending each other.

\begin{proposition} \label{proplgmod4} 
Let $P_1, P_2$ be two programs such that $P_1$ extends $P_2$ and 
$P_2$ extends $P_1$.  Let $S \subseteq B^E_{P_1 \cup P_2}$. If
\begin{itemize}
\item $P_1$ LG-terminates w.r.t.\ $Tab_{P_1}$ and $S \cap B_{P_1}^E$,
\item $P_2$ LG-terminates w.r.t.\ $Tab_{P_2}$ and $S \cap B_{P_2}^E$,
\end{itemize}
then $P_1 \cup P_2$ LG-terminates w.r.t.\ $Tab_{P_1\cup P_2}$ and $S$.
\end{proposition}

\begin{proof}
Because $P_1$ extends $P_2$ and $P_2$ extends $P_1$, 
$Call(P_1\cup P_2, S) \cap B^E_{P_i} = Call(P_i, S\cap B^E_{P_i})$, 
for $i =1,2$.   
The proposition follows then by definition of LG-termination.
\qed
\end{proof}

Similar as in the case of quasi-termination, the above modular,
sufficient conditions for the LG-termination of $P\cup R$ are in many
cases also necessary.  In particular, the modular conditions of
Proposition~\ref{propmodlg} are also necessary for the LG-termination
of $P \cup R$ in case the tabling $Tab_P$ is well-chosen w.r.t.~$P$.
Because in all cases where $Def_P \subseteq NTab_P$, respectively
$Def_P \subseteq Tab_P$, $Tab_P$ is well-chosen w.r.t.~$P$, it
follows that the modular conditions of Proposition~\ref{proplgmod2},
respectively Proposition~\ref{proplgmod3}, are necessary in general.
Also the modular conditions of Propositions~\ref{proplgmod4} are
necessary in general.

\subsection{Construction of Level Mappings in Modular Termination Proofs}
\label{ssectconstrlm}

We now take a closer look at the modular termination proofs of the
previous subsections.  We follow the approach of \cite{apt:modular},
where modular proofs for SLD-termination (i.e.\ termination of
SLD-resolution w.r.t.\ all selection rules) and LD-termination are
given.  In \cite{apt:modular}, (S)LD-termination of a program $P\cup
R$, where $P$ extends $R$, is proven by constructing a level mapping
$\level{.}$ for $P\cup R$ which satisfies some acceptability
condition.  The level mapping $\level{.}$ is constructed from simpler
level mappings for the separate components $P$ and $R$.  Namely,
$\level{.}$ is constructed from $\level{.}_P$, $\level{.}_R$ and
$\llevel{.}_P$, where $\level{.}_P$, respectively~$\level{.}_R$, is a
level mapping for $P$, respectively~$R$, satisfying the acceptability
condition, and where $\llevel{.}_P$ is a level mapping for $P$ serving
as the connecting part between the two components.  The level mapping
$\level{.}$ for $P\cup R$ is then defined as $\level{.}_P +
\llevel{.}_P$ on the atoms defined in $P$ and as $\level{.}_R$ on the
atoms defined in $R$.  It is proven that such a construction always
returns a level mapping satisfying the acceptability condition for the
whole program $P\cup R$.

We follow the same approach for the case of quasi-termination (we do
not consider LG-termination since it can be dealt with in a similar
way).  In particular, we give modular proofs of the quasi-termination
of a program $P \cup R$, where $P$ extends $R$, by constructing a
level mapping such that $P \cup R$ is quasi-acceptable w.r.t.\/ this
level mapping (see Definition~\ref{defquasiacce} and
Theorem~\ref{theoquasichar}).  The construction of such a level
mapping is done in a way similar to~\cite{apt:modular}, which we
explained above.  We first need the following lemma, which gives
sufficient, modular conditions on a level mapping in order to be
finitely partitioning on some subset of the extended Herbrand base.
We use a slightly more general definition of a level mapping, namely,
a level mapping is a mapping from a \emph{subset} of the extended
Herbrand base to the natural numbers.

\begin{lemma} \label{lem2} \hfill
\begin{enumerate}
\item Let $P$ be a program and $L \subseteq B_P^E$.
      Let $\level{.}, \llevel{.}: L \ra \bbbn$ be level mappings.
      If $\level{.}$ is finitely partitioning on $C \subseteq L$, then
      $\level{.} + \llevel{.} : 
        L \ra \bbbn : A \mapsto (\level{.}+\llevel{.})(A) =
                                 \level{A} + \llevel{A}$ 
      is finitely partitioning on $C$.

\item Let $P_1, P_2$ be two programs and $L_1 \subseteq B_{P_1}^E$, 
      $L_2 \subseteq B_{P_2}^E$.  Let $\level{.}_1: L_1 \ra \bbbn$ and
      $\level{.}_2: L_2 \ra \bbbn$ be level mappings.
      If $\level{.}_1$, respectively $\level{.}_2$, is finitely 
      partitioning on $C_1\subseteq L_1$, respectively
      $C_2 \subseteq L_2$, then 
      $m(\level{.}_1,\level{.}_2): L_1 \cup L_2 \ra \bbbn:$ 
      \[
        A \mapsto m(\level{.}_1,\level{.}_2)(A) := 
        \left\{
            \begin{array}{lll}
            min(\level{A}_1, \level{A}_2) & , & A \in L_1 \cap L_2 \\
                \level{A}_1 & , & A \in L_1 \setminus L_2 \\
                \level{A}_2 & , & A \in L_2 \setminus L_1
            \end{array}
        \right.
      \]
      is finitely partitioning on $C_1 \cup C_2$.
\end{enumerate}
\end{lemma}

\begin{proof} \hfill

\begin{enumerate}
\item \label{itemfinpart1}
Let $n \in \bbbn$. We prove that 
$\sharp((\level{.}+\llevel{.})^{-1}(n)\cap C) < \infty$.
\[
\begin{array}{lll}
(\level{.}+\llevel{.})^{-1}(n) \cap C & = & 
\{ A \in C~|~(\level{.}+\llevel{.})(A) = n \}
\\
 & \subseteq & \{ A \in C~|~\level{A} \leq n \}
\\
 & = & \bigcup_{0 \leq m \leq n} \{ A \in C~|~\level{A} = m \} 
\end{array}
\]
and this last set is finite.

\item \label{itemfinpart2}
Let $n \in \bbbn$. We prove that 
$\sharp (m(\level{.}_1,\level{.}_2)^{-1}(n)$
$\cap$ $(C_1 \cup C_2)) <$ $\infty$.
\[
\begin{array}{lll}
m(\level{.}_1,\level{.}_2)^{-1}(n) \cap (C_1 \cup C_2) &
= & \{ A \in C_1 \cup C_2~|~m(\level{.}_1,\level{.}_2)(A) = n \} 
\\
 & = & \{ A \in C_1 \setminus C_2~|~\level{A}_1 = n \} \cup
\\
 & & \{ A \in C_2 \setminus C_2~|~\level{A}_2 = n \} \cup
\\
 & & \{ A \in C_1 \cap C_2~|~min(\level{A}_1,\level{A}_2) = n \} 
\end{array}
\]
It is obvious that the first two sets in the union are finite 
($\level{.}_1$, resp.\ $\level{.}_2$, 
is finitely partitioning on $C_1$, resp.\ $C_2$).
The set 
$\{ A \in C_1 \cap C_2~|~min(\level{A}_1,\level{A}_2)$ $= n \}$ is 
finite, because
it is a subset of the finite set
$\{ A \in C_1 \cap C_2~|~\level{A}_1= n \}$
$\cup$
$\{ A \in C_1 \cap C_2~|~\level{A}_2= n \}$.
\end{enumerate}
\qed \end{proof}

We next give a modular termination condition for the quasi-termination
of $P\cup R$ where $P$ extends $R$, by constructing a level mapping,
from simpler ones, such that $P\cup R$ is quasi-acceptable w.r.t.\
this level mapping.  Notice that we consider the same case as in
Proposition~\ref{propqmod1}.

\begin{proposition} \label{propconstrlm}
Let $P$ and $R$ be two programs such that $P$ extends $R$ and
let $S \subseteq B_{P\cup R}^E$.
If

\begin{enumerate}

\item \label{it1}
$R$ is quasi-acceptable w.r.t.\ $Tab_R$, $Call(P\cup R,S)$ and
the level mapping $\level{.}_R$, 
defined on $B_R^E$ and finitely partitioning on 
$Call(P\cup R,S) \cap B_{Tab_R}^E$,

\item \label{it2}
there is a level mapping $\level{.}_P$ defined on $B_P^E \setminus B_R^E$
and finitely partitioning on 
$Call(P\cup R,S) \cap B^E_{Tab_P \setminus Tab_R}$
such that
\begin{itemize}

\item
for every atom $A$ such that $\tilde{A} \in Call(P\cup R,S)$,
\item
for every clause $H \la B_1,\ldots,B_n$ in $P$, such that 
$mgu(A,H) = \theta$ exists,
\item
for every $1 \leq i \leq n$ such that
$B_i \in B_P^E \setminus B_R^E$,

\item
for every $cas$ $\theta_{i-1}$ in 
$P\cup R$ for $\la (B_1,\ldots,B_{i-1})\theta$:
\[
\begin{array}{ll}
\level{A}_P \geq  \level{B_i \theta \theta_{i-1}}_P &
\\
\mbox{and} &
\\
\level{A}_P >  \level{B_i \theta \theta_{i-1}}_P & \mbox{if}~Rel(A) 
\simeq Rel(B_i) \in NTab_P~\mbox{and}
\\
 & C_2(Rel(A),Rel(B_i))~\mbox{does not hold} .
\end{array}
\]
\end{itemize}

\item \label{it3}
there exists a level mapping $\llevel{.}_P$ on 
$B^E_P \setminus B^E_R$ such that

\begin{itemize}
\item
for every atom $A$ such that $\tilde{A} \in Call(P \cup R,S)$,
\item
for every clause $H \la B_1, \ldots, B_n$ in $P$, such that $mgu(A,H)= \theta$
exists,
\item
for every $1 \leq i \leq n$,

\item
for every $cas$ $\theta_{i-1}$ in $P \cup R$ 
for $\la (B_1, \ldots, B_{i-1})\theta$:
\end{itemize}
\[
\llevel{A}_P \geq 
\left\{
\begin{array}{lll}
\llevel{B_i \theta \theta_{i-1}}_P & , 
& B_i \theta \theta_{i-1}
\in B^E_P \setminus B^E_R
\\
\level{B_i \theta \theta_{i-1}}_R & , & B_i \theta \theta_{i-1} \in B^E_R
\end{array}
\right.
\]

\end{enumerate}
then, the following level mapping 
$\level{.}$, defined on $B_{P\cup R}^E$, is finitely partitioning on 
$Call(P\cup R,S) \cap B^E_{Tab_{P\cup R}}$:
\[
\level{A} = 
\left\{
\begin{array}{ll}
\level{A}_P + \llevel{A}_P & \mbox{if}~A \in B_P^E \setminus B_R^E,
\\
\level{A}_R & \mbox{if}~A \in B_R^E,
\end{array}
\right.
\]
and $P \cup R$ is quasi-acceptable w.r.t.\ $Tab_{P\cup R}$,
$S$, and the level mapping $\level{.}$.
Hence, $P\cup R$ quasi-terminates w.r.t.\ $Tab_{P\cup R}$ and $S$.
\end{proposition}

\begin{proof}
Because of Lemma~\ref{lem2}(\ref{itemfinpart1},\ref{itemfinpart2}), 
the level mapping $\level{.}$ is finitely
partitioning on $Call(P\cup R,S) \cap B^E_{Tab_{P\cup R}}$.
We prove that $P\cup R$ is quasi-acceptable w.r.t.\ $Tab_{P\cup R}$,
$S$ and the level mapping $\level{.}$ (see Definition~\ref{defquasiacce}).
\\ 
Let $A$ be an atom such that $\tilde{A} \in Call(P \cup R,S)$.
Let $H \la B_1, \ldots, B_n$ be a clause of $P \cup R$ such that 
$mgu(A,H) = \theta$ exists.
Let $\theta_{i-1}$ be a $cas$ in $P \cup R$ for
$\la (B_1, \ldots, B_{i-1})\theta$. 
There are two cases to consider:
\begin{itemize}
\item
$A$ is defined in $R$, $\level{A} = \level{A}_R$.

Then, because of condition \ref{it1} in the proposition statement,
$\level{A}_R \geq \level{B_i \theta \theta_{i-1}}_R$ 
(note that because $P$ extends $R$, 
for a clause $H \la B_1, \ldots, B_n$ in $R$
and $mgu(A,H) = \theta$, a $cas$ for 
$\la (B_1, \ldots, B_{i-1})\theta$ in $P \cup R$ is the same as a 
$cas$ for 
$\la (B_1, \ldots, B_{i-1})\theta$ in $R$ only).
Since $A, B_i \theta \theta_{i-1}$ $\in B_R^E$, 
$\level{A} = \level{A}_R \geq$
$\level{B_i \theta \theta_{i-1}}_R = \level{B_i \theta \theta_{i-1}}$.
In case $Rel(A) \simeq Rel(B_i) \in NTab_R$ and 
$C_2(Rel(A),Rel(B_i))$ does not hold, $\level{A} = \level{A}_R >$
$\level{B_i \theta \theta_{i-1}}_R = \level{B_i \theta \theta_{i-1}}$.

\item
$A$ is defined in $P$, $\level{A} = \level{A}_P + \llevel{A}_P$.

\begin{itemize}

\item
$B_i \in B_R^E$, $\level{B_i \theta \theta_{i-1}} = 
\level{B_i \theta \theta_{i-1}}_R$.

Because of condition \ref{it3} in the proposition statement,
$\llevel{A}_P$ $\geq$ $\level{B_i \theta \theta_{i-1}}_R$.
Hence, $\level{A}=$ 
$\level{A}_P + \llevel{A}_P$ $\geq$ $\level{B_i \theta \theta_{i-1}}_R$
$=\level{B_i \theta \theta_{i-1}}$.

Note that in this case we always have that $Rel(A) \not\simeq Rel(B_i)$
(because $P$ extends $R$).

\item
$B_i \in B_P^E \setminus B_R^E$, $\level{B_i \theta \theta_{i-1}} =$
$\level{B_i \theta \theta_{i-1}}_P + $
$\llevel{B_i \theta \theta_{i-1}}_P$.

Because of condition \ref{it2} in the proposition statement,
$\level{A}_P$ $\geq$ $\level{B_i \theta \theta_{i-1}}_P$.  Also, because of 
condition \ref{it3}, 
$\llevel{A}_P$ $\geq$ 
$\llevel{B_i \theta \theta_{i-1}}_P$.
Hence,
$\level{A}_P +$ $\llevel{A}_P$ $\geq$ 
$\level{B_i \theta \theta_{i-1}}_P +$
$\llevel{B_i \theta \theta_{i-1}}_P$.
In case $Rel(A) \simeq Rel(B_i) \in NTab_P$ and 
$C_2(Rel(A),Rel(B_i))$ does not hold,
we have that $\level{A}_P$ $>$ $\level{B_i \theta \theta_{i-1}}_P$,
hence
$\level{A}_P + \llevel{A}_P$ $>$ 
$\level{B_i \theta \theta_{i-1}}_P + $
$\llevel{B_i \theta \theta_{i-1}}_P$.
\end{itemize}
\end{itemize}
In each case, we conclude that $\level{A}$ $\geq$ 
$\level{B_i \theta \theta_{i-1}}$ and that, in case
$Rel(A) \simeq Rel(B_i) \in NTab_{P\cup R}$ and 
$C_2(Rel(A),Rel(B_i))$ does not hold,
$\level{A}$ $>$ 
$\level{B_i \theta \theta_{i-1}}$.
\qed \end{proof}

\begin{example} \label{exaconstrlm}
Recall the program $P\cup R$ of Example~\ref{examodular} 
(see also Examples~\ref{exaqmod3} and \ref{exapath}).
\[
\begin{array}{ll}
P: & \left\{
        \begin{array}{lll}
          path(X,Ed,Y,[Y]) & \la & edge(X,Ed,Y) \\
          path(X,Ed,Z,[Y|L]) & \la & edge(X,Ed,Y),\ path(Y,Ed,Z,L)
        \end{array}
     \right.
\\
~
\\
R: & \left\{
        \begin{array}{lll}
          edge(X,[e(X,Y)|L],Y) & \la & \\
          edge(X,[e(X_1,X_2)|L],Y) & \la & edge(X,L,Y)
        \end{array}
     \right.
\end{array}
\]
Let $Tab_{P\cup R} = \{\mathit{path}/4\}$ and $S =
\{\mathit{path(a,[e(a,b),e(b,a)],Y,L)}\}$.  We prove that $P\cup R$
quasi-terminates w.r.t.\ $\{\mathit{path}/4\}$ and $S$ using the above
proposition.  The first two conditions of this proposition were
already tackled in Example~\ref{exaqmod3} (using a different set $S$;
the arguments remain the same however).
\begin{enumerate}
\item $R$ is quasi-acceptable w.r.t.\ $\emptyset$, $Call(P\cup R,S)$
      and the level mapping $\level{.}_R$:
      \[
        \level{edge(t_1,t_2,t_3)}_R = \norm{t_2}_l .
      \]
\item The trivial level mapping, $\level{.}_P$, on $B^E_P \setminus
      B^E_R = B^E_{\{\mathit{path}\}}$ satisfies the second condition
      of the proposition; $\mathit{path}/4$ is tabled so a strict
      decrease is never required and there are a finite number of
      $\mathit{path}$-atoms in the call set.
\item The following level mapping, $\norm{.}_P$, on $B^E_P \setminus
      B^E_R = B^E_{\{\mathit{path}\}}$ satisfies the third condition
      of the proposition:
      \[
        \norm{path(t_1,t_2,t_3,t_4)}_P = \norm{t_2}_l .
      \]
\end{enumerate}
Hence, the level mapping $\level{.}$, on $B^E_{P\cup R}$,
is defined as follows:
\[
  \begin{array}{lll}
        \level{path(t_1,t_2,t_3,t_4)} & = & \norm{t_2}_l \\
        \level{edge(t_1,t_2,t_3)} & = & \norm{t_2}_l 
  \end{array} 
\]
and by the proposition, 
the program 
$P \cup R$ is quasi-acceptable w.r.t.\ $Tab_{P\cup R} =
\{\mathit{path}/4\}$, $S = \{\mathit{path(a,[e(a,b),e(b,a)],Y,L)}\}$
and the level mapping $\level{.}$.
\end{example}

In~\cite{apt:pedreschi} the special case where two programs use
disjoint sets of predicates is also considered.  We next consider a
more general case in which the two programs may use the same
predicates but may not define the same predicate; that is, both
programs extend each other.  This case was already considered in
Proposition~\ref{propqmod4}.

\begin{proposition} \label{propnext}
Let $P_1, P_2$ be programs such that $P_1$ extends $P_2$ and $P_2$
extends $P_1$.  Let $S \subseteq B^E_{P_1 \cup P_2}$.  Suppose that
\begin{enumerate}
\item \label{propita}
        $P_1$ is quasi-acceptable w.r.t.\/ $Tab_{P_1}$, $S \cap
        B_{P_1}^E$ and a level mapping $\level{.}_1$ on $B_{P_1}^E$
        which is finitely partitioning on
        $Call(P_1, S \cap B_{P_1}^E) \cap B^E_{Tab_{P_1}}$,
\item \label{propitb}
        $P_2$ is quasi-acceptable w.r.t.\/ $Tab_{P_2}$, $S \cap
        B_{P_2}^E$ and a level mapping $\level{.}_2$ on $B_{P_2}^E$
        which is finitely partitioning on
        $Call(P_2, S \cap B_{P_2}^E) \cap B^E_{Tab_{P_2}}$,
\end{enumerate}
then $m(\level{.}_1,\level{.}_2)$ (see Lemma~\ref{lem2},
point~\ref{itemfinpart2}) is a finitely partitioning level mapping on
$Call(P_1 \cup P_2,S) \cap B^E_{Tab_{P_1 \cup P_2}}$, and $P_1 \cup
P_2$ is quasi-acceptable w.r.t.\ $Tab_{P_1 \cup P_2}$, $S$ and
$m(\level{.}_1,\level{.}_2)$.  Hence, $P_1 \cup P_2$ quasi-terminates
w.r.t.\ $Tab_{P_1 \cup P_2}$ and $S$.
\end{proposition}

\begin{proof}
Note that because $P_1$ extends $P_2$ and vice versa,
$Call(P_1 \cup P_2,S) \cap B_{P_i}^E$ 
$=$ $Call(P_i, S \cap B_{P_i}^E)$, $i \in \{1,2\}$.
By Lemma~\ref{lem2}(\ref{itemfinpart2}), 
$m(\level{.}_1,\level{.}_2)$ is finitely partitioning on 
$Call(P_1 \cup P_2,S) \cap B^E_{Tab_{P_1 \cup P_2}}$.
Also, if 
$H \la B_1, \ldots, B_n$ is a clause in $P_i$ and $mgu(A, H)= \theta$, 
then a $cas$ in $P_i$ for 
$\la(B_1, \ldots, B_{i-1})\theta$ is
a $cas$ in $P_i \cup P_j$ ($\{i,j\} = \{1,2\}$) 
for $\la(B_1, \ldots, B_{i-1})\theta$ and vice versa.
Then it directly follows that $P_1 \cup P_2$ is quasi-acceptable w.r.t.\
$Tab_{P_1 \cup P_2}$, $S$ and $m(\level{.}_1,\level{.}_2)$.
\qed
\end{proof}

As was shown in~\cite{apt:modular} for the case of (S)LD-termination,
the above modular conditions provide us with an incremental, bottom up
method for proving termination of tabled logic programs.

\section{Towards Automated Termination Proofs for Tabled Logic Programs}
\label{sectauto}

Having described the basic framework for proving termination of tabled
logic programs, in this section we examine issues related to the {\em
automation} of the termination conditions.  We will only consider
quasi-termination in this section; the results for LG-termination
carry over in the same way.  We show how to extend the
constraint-based, automatic approach for proving LD-termination of
Decorte, De Schreye and Vandecasteele~\cite{stef:constraints}, in
order to prove quasi-termination
of tabled logic programs
in an automatic way.  Our results hold for the class of simply moded,
well-moded programs and queries.

We first recall the main ideas of \cite{stef:constraints}.  In
\cite{stef:constraints}, a new strategy for automatically proving
LD-termination of logic programs w.r.t.\/ sets of queries is
developed.  A symbolic termination condition is introduced, called
\emph{rigid acceptability}, by parametrising the concepts of norm,
level mapping and model.  The rigid acceptability condition is
translated into a system of constraints on the values of the
introduced symbols only.  A system of constraints identifies sets of
suitable norms, level mappings and models which can be used in the
termination condition. In other words, if a solution for the
constraint system exists, termination can be proved.  The solving of
constraint sets enables the different components of a termination
proof to communicate with one another and to direct the proof towards
success (if there is).  The method of \cite{stef:constraints} is both
efficient and precise.

This section\footnote{For the referees, we want to mention that the
material presented in this section is very related to results
presented in the article {\em ``Termination of simply moded well-typed
programs under a tabled execution mechanism''} (S. Verbaeten and D. De
Schreye), which is submitted to the Journal of Applicable Algebra in
Engineering, Communication and Computing (AAECC).  There, we present
similar results for the special case in which all the predicates of
the program are tabled.  Here we extend those results to allow a
mix of tabled and non-tabled predicates.  In spite of the partial
overlap with this article, we choose to include this section anyway,
because we feel that it is important to explicitly discuss the
prospects of automating the results presented in the previous
sections.  In the article submitted to AAECC, we consider the larger
class of simply moded well-typed programs, instead of simply moded
well-moded programs that we consider here.  With respect to the
automation of the method, this larger class adds nothing new, so we do
not include these results here.  If the referees think this section
should not be included, we are willing to remove it.}
is structured as follows.  We first reformulate the
quasi-accep\-ta\-bi\-li\-ty condition into a condition at the clause level,
which is needed in the constraint-based termination analysis framework
of~\cite{stef:constraints}.  This gives us the rigid
quasi-acceptability condition.  Then, we recall the symbolic forms for
norm, level mapping and model, as introduced
in~\cite{stef:constraints}.  We introduce the class of simply moded,
well-moded programs and queries, for which we translate the rigid
quasi-accep\-ta\-bi\-li\-ty condition into a system of constraints
on the introduced symbols.

\subsection{Rigid Quasi-Acceptability Condition} \label{ssectrigidacce}

In order to prove termination in 
an automatic, constraint-based way
as in~\cite{stef:constraints}, it is important to have a termination
condition which is stated at the clause level (and not on 
sets of calls
as 
the quasi-acceptability condition of Theorem~\ref{theoquasichar} is).
In most automatic approaches, and in particular in that
of~\cite{stef:constraints}, this is obtained by requiring that the
level mapping is \emph{rigid} on the call set.  A level mapping is
rigid on the call set iff the value of an atom in the call set is
invariant under substitutions.  If a level mapping is rigid on the
call set, the atoms in the call set can be considered as ground
w.r.t.\ the level mapping.  In this way, the problem of
back-propagation of bindings in the calls is dealt with, and 
this allows
the termination condition 
to
be stated at the clause level (see also~\cite{survey}).

\begin{definition}[rigid level mapping]
Let $P$ be a program and $C \subseteq B_{P}^E$.
A level mapping $\level{.}$ is {\em rigid on $C$}
iff for all atoms $A \in C$, for all substitutions $\psi$,
$\level{A} = \level{A\psi}$.
\end{definition}

The following condition of rigid quasi-acceptability is derived from
the quasi-acceptability condition, and will serve as the basis for a
constraint-based, symbolic condition for quasi-termination.

\begin{proposition}[rigid quasi-acceptability condition] \label{propquswm}
Let $P$ be definite program, $Tab_P \subseteq Pred_P$ and 
$S \subseteq B_P^E$ be a set of queries.
\\
If there exists a level mapping $\level{.}$,
such that $\level{.}$ is rigid on $Call(P,S)$,
$\level{.}$ is finitely partitioning on $Call(P,S) \cap B^E_{Tab_P}$,
and such that 
\begin{itemize}

\item
for every clause $H \la B_1,\ldots,B_n$ in $P$,

\item
for every atom $B_i$, $i \in \{1,..,n\}$,

\item
for every 
substitution $\psi$ 
such that 
$P\models B_1\psi, \ldots, B_{i-1}\psi$:

\[
\begin{array}{ll}
\level{H \psi} \geq \level{B_i \psi} &
\\
\mbox{and} &
\\
\level{H\psi} > \level{B_i\psi} & \mbox{if}~ 
Rel(H) \simeq Rel(B_i) \in NTab_P~\mbox{and}
\\
 & C_2(Rel(A),Rel(B_i))~\mbox{does not hold},
\end{array}
\] 
\end{itemize}
then $P$ is quasi-acceptable w.r.t.\ $Tab_P$ and $S$.
Hence, $P$ quasi-terminates w.r.t.\ $Tab_P$ and $S$.
\end{proposition}

\begin{proof}
Suppose the above condition is satisfied for $P$.
We prove that $P$ is quasi-acceptable w.r.t.\ $Tab_P$, $S$ and the
level mapping $\level{.}$.
Let $A$ be an atom such that $\tilde{A} \in Call(P,S)$.
Let  
$H \la B_1,\ldots,B_n$ be a clause 
in $P$ such that $mgu(A,H)=\theta$ exists.
Let $\theta_{i-1}$ be an LD- computed answer substitution for 
$\la (B_1,\ldots,B_{i-1})\theta$.
Then, 
$P\models B_1\theta\theta_{i-1}, \ldots, B_{i-1}\theta\theta_{i-1}$.
We prove that $\level{A} \geq \level{B_i\theta\theta_{i-1}}$.
By the condition in the proposition, we know that 
$\level{H\theta\theta_{i-1}} \geq \level{B\theta\theta_{i-1}}$.
Because $A\theta = H\theta$, 
$\level{A\theta\theta_{i-1}} = \level{H\theta\theta_{i-1}}$.
Now, since $A \in Call(P,S)$ and $\level{.}$ is rigid on 
$Call(P,S)$, $\level{A\theta\theta_{i-1}} = \level{A}$.
Thus, $\level{A} = \level{H\theta\theta_{i-1}}$,
and
therefore $\level{A} \geq \level{B_i\theta\theta_{i-1}}$.
The proof that $\level{A} > \level{B_i\theta\theta_{i-1}}$
in case $Rel(A) \simeq Rel(B_i) \in NTab_P$ and
$C_2(Rel(A),Rel(B_i))$ does not hold, is analogous.
\qed
\end{proof}

\subsection{Symbolising the Concepts of Norm, Level Mapping and Model}
\label{ssectsymbols}

We recall the symbolic forms for norms, level mappings
and interargument relations (which are abstractions of models),
as introduced in \cite{stef:constraints}.
These will form the basis for the symbolic termination condition.
The condition will be formulated as a search for suitable values for all
introduced symbols.
We refer to \cite{stef:constraints} for motivation and more details.

We will need the following notions.

\begin{definition}[functor, predicate and extended predicate 
coefficients]
The set of 
{\em functor coefficients}, respectively 
{\em predicate coefficients},
respectively {\em extended predicate coefficients} associated to a 
program $P$ are the sets of symbols
\[
\begin{array}{l}
  FC(P) = \{ f_i\ |\ f/n \in Fun_P \wedge i \in \{0,...,n\}\} ,  \\
  PC(P) = \{ p_i\ |\ p/n \in Pred_P \wedge i \in \{1,...,n\}\} , \\
  EC(P) = \{ p_i^e\ |\ p/n \in Pred_P \wedge i \in \{0,...,n\}\} .
\end{array}
\]
\end{definition}

Let ${\cal C}$ denote the set of symbols $FC(P) \cup PC(P) \cup
EC(P)$.  The symbol $L_{<{\cal C};+,.;\leq>}$ denotes the language
containing the symbols in the set ${\cal C}$ as constants, the infix
functor $+/2$, the infix functor $./2$, the relation symbol $\leq/2$
and the set of variables in the first order language of the program
$P$.  Terms in that language are defined in the usual way.  The
relation symbols $=/2$ and $</2$ are defined in terms of $\leq/2$ as
usual and considered as additional primitive predicates.  We denote
the set of all possible atoms in that language by $S_{<{\cal
C};+,.;\leq>}$.  We call such atoms symbolic expressions.  The set of
all logical formulae over $S_{<{\cal C};+,.;\leq>}$ is denoted as
$F_{<{\cal C};+,.;\leq>}$.  Such formulae are called symbolic
formulae.  By natural formulae, we denote formulae in
$F_{<\{0,1\};+,.;\leq>}$.  For a formula $F$, $\forall F$ denotes the
universal closure over the free variables occurring in $F$.

We introduce symbolic (semi-linear) norms.

\begin{definition}[symbolic norm $\norm{.}^s$]
Let $FC(P)$ be a set of functor coefficients.
\[
\begin{array}{lclcl}
\norm{.}^s: & Term_P & \rightarrow & S_{<{\cal C};+,.;\leq>} & \\
 & t & \rightarrow & f_0 + \sum_{i=1}^nf_i\norm{t_i}^s & 
                \mbox{ if } t = f(t_1,\cdots,t_n), n > 0, \\
 & t & \rightarrow & 0 & \mbox{ if } t = c \in Const_P, \\
 & t & \rightarrow & X & \mbox{ if } t = X \mbox{ is a variable}.
\end{array}
\]
\end{definition}

Note that the symbolic norm of a variable is the variable itself.
So, the symbolic norm includes information about the instantiation
level of the term. That is, the symbolic norm takes into account those
parts of the concrete term whose size may still change under instantiation.

In the same way, we can define a symbolic level mapping by symbolising 
its coefficients.

\begin{definition}[symbolic level mapping $\level{.}^s$]
Let $PC(P)$ be a set of predicate coefficients and $\norm{.}^s$
a symbolic norm.
\[
\begin{array}{lcll}
\level{.}^s : & Atom_P & \rightarrow & S_{<{\cal C};+,.;\leq>}  \\
& p(t_1,\cdots,t_n) & \rightarrow & \sum_{i=1}^np_i\norm{t_i}^s . \\
\end{array}
\]
\end{definition}

Finally, we want to abstract the notion of model.
Norms allow to abstract models by specifying relations which hold between
the size of certain arguments of their member atoms.
This leads to the notion of interargument relation.

\begin{definition}[(valid) interargument relation]
Let $p/n \in Pred_P$. 
\\
An {\em interargument relation for $p/n$}
is a relation $R^{p/n} \subseteq \bbbn^n$.
\\
An interargument relation $R^{p/n}$ for $p/n$ is a 
{\em valid interargument relation w.r.t.\ a norm \norm{.}}
iff $\forall p(t_1,\ldots,t_n) \in Atom_P$:
if $P \models p(t_1,\ldots,t_n)$, then
$(\norm{t_1},\ldots,\norm{t_n})$ $\in R^{p/n}$.
\end{definition}
 
As 
in~\cite{stef:constraints}, we will allow interargument relations
which express an inequality relation: $R^{p/n} =
\{(x_1,\ldots,x_n)~|~\sum_{i \in I_p}k_i x_i \geq \sum_{j \in O_p}k_j
x_j + k_0\}$, with $k_i \in \bbbn$, $i \in \{0,1,\ldots,n\}$,
depending on $p/n$, $I_p \cup O_p \subseteq \{1,\cdots,n\}$ and $I_p
\cap O_p = \emptyset$.  The sets $I_p$ and $O_p$ are assumed fixed for
each predicate.  In \cite{stef:constraints}, it is argued that these
sets can best be seen as some kind of a generalisation of the sets of
input and output arguments.  In the following subsection, we will
assume a fixed mode for each predicate, and then the set $I_p$,
respectively $O_p$, will be taken as the set of input, respectively
output, positions of the predicate $p$.

Finally, we can abstract success sets by abstracting interargument
relations.

\begin{definition}[symbolic size expression ${\cal A}^s$]
\label{symbolic_sizeexp}
Let $EC(P)$ be a set of extended predicate coefficients and
$\norm{.}^s$ a symbolic norm.
\[
\begin{array}{lcll}
 {\cal A}^s: & Atom_P & \rightarrow & S_{<{\cal C};+;\leq>}  \\
 & t_1 = t_2 & \rightarrow & \norm{t_1}^s = \norm{t_2}^s , \\
 & p(t_1,\cdots,t_n) & \rightarrow & \sum_{i \in I_p}p_i^e\norm{t_i}^s \geq 
\sum_{j \in O_p}p_j^e\norm{t_j}^s + p_0^e, \mbox{ where }p \not= = .
\end{array}
\]
\end{definition}

\begin{definition}[symbol mapping]
A {\em symbol mapping} is a mapping $s: {\cal C} \rightarrow \bbbn$.
\end{definition}

Expressions involving only symbols from ${\cal C}$ are mapped into the
natural numbers by substituting the symbols by their mapped value.
With abuse of notation, if $F$ is a symbolic formula and $s$ a 
symbol mapping, we denote the associated natural formula as
$s(F)$.

Each symbol mapping induces in a natural way a norm, level mapping and
interargument relations.  More precisely, the symbol mapping $s$
induces the following norm $\norm{.}_s$:
\[
\left\{
\begin{array}{ll}
\norm{X}_s = \norm{c}_s = 0 & \mbox{if}~X~\mbox{is a variable}, c \in Const_P,
\\
\norm{f(t_1,\ldots,t_n)}_s = s(f_0) + \sum_{i=1}^n s(f_i) \norm{t_i}_s , &
\end{array}
\right.
\]
level mapping $\level{.}_s$:
\[
\level{p(t_1,\ldots,t_n)}_s = \sum_{i=1}^n s(p_i) \norm{t_i}_s ,
\]
and interargument relations $R^{p/n}_s$:
\[
\{(\norm{t_1}_s) , \ldots, \norm{t_n}_s) \ |\ 
t_1,\ldots,t_n \in Term_P\ \mbox{and}\ s ( {\cal A}^s(p(t_1,\ldots,t_n))) 
\mbox{ holds } \}.
\]

Our aim is to formulate the rigid quasi-acceptability condition
in a constraint-based way.
In particular this means that we have to find syntactical conditions
on a symbol mapping $s$ such that
\begin{itemize}
\item
$\level{.}_s$ is rigid on $Call(P,S)$, and
\item
$\level{.}_s$ is finitely partitioning on $Call(P,S) \cap B^E_{Tab_P}$.
\end{itemize}
As
shown in e.g.~\cite{bossi:tapsoft91}, a level mapping
$\level{.}_s$ is rigid on the call set if
it does {\em not} take into account {\em too many} argument positions
of predicates and functors in its linear combination\footnote{More 
precisely,
$s$ is equal to $0$ on all functor and predicate
coefficients for which there is an atom in the call set which contains 
a variable on such a position. For a more formal discussion on 
this topic, we refer to \cite{VerbaetenCW99,Verbaeten_flops99}.}. 
On the other hand,
a level mapping $\level{.}_s$ is 
finitely partitioning on $Call(P,S) \cap B^E_{Tab_P}$, 
if {\em enough} argument positions
are taken into account in its linear combination\footnote{That is, the
symbol mapping $s$ is different from $0$ on enough functor and 
predicate coefficients. For a more formal discussion on 
this topic, we again refer to \cite{VerbaetenCW99,Verbaeten_flops99}.}.
We
will
show that, for the class of simply moded, well-moded programs and
queries, we are able to combine these two---at first sight 
contradictory---
conditions.  We first introduce the class of simply moded, well-moded
programs and queries and then formulate, for this class, the symbolic
termination condition.

\subsection{The Class of Simply Moded, Well-Moded Programs and Queries}
\label{ssectsmwm}

\begin{definition}[mode for a predicate]
Let $p$ be an $n$-ary predicate symbol.  A {\em mode} for $p$ is a
function $m_p:$ $\{1,\ldots,n\} \ra$ $\{In,Out\}$.  If $m_p(i) = In$
(respectively $Out$), then we say that $i$ is an {\em input}
(respectively {\em output}) {\em position} of $p$ (w.r.t.\ $m_p$).
\end{definition}

We assume that each predicate symbol has a unique mode.
For predicates that have multiple modes, we can assume that a
straightforward renaming process taking each mode into account has
already been performed.
In examples, we will write the mode $m_p$ for the predicate $p$
as follows: $p(m_p(1),\ldots,m_p(n))$.
Given a predicate $p \in Pred_P$ with mode $m_p$, we denote by
$I_p = \{i~|~m_p(i) = In\} $ the set of input positions of $p$ according
to $m_p$, and by 
$O_p = \{i~|~m_p(i) = Out\} $ the set of output positions of $p$ according
to $m_p$.

We recall the notion of well-modedness (see e.g.\ \cite{apt:emarchiori}).
For simplifying the notation, when writing an atom as $p({\bf u},{\bf v})$,
we assume that ${\bf u}$ is the sequence of terms filling in the input
positions of $p$ and ${\bf v}$ is the sequence of terms filling in the 
output positions of $p$. 
For a term $t$, we denote by $Var(t)$ the set of variables occurring in $t$.
Similar notation is used for sequences of terms.

\begin{definition}[well-modedness] \label{defwm}
~\\
A clause $p_0({\bf t_0},{\bf s_{n+1}}) \la$ 
$p_1({\bf s_1},{\bf t_1}),\ldots,p_n({\bf s_n},{\bf t_n})$ is
called {\em well-moded} iff for $i \in [1,n+1]$:
\[
Var({\bf s_i}) \subseteq \bigcup_{j=0}^{i-1} Var({\bf t_j}) .
\]
A program is called {\em well-moded} iff every clause of it is well-moded.
\\
A query 
$\la p_1({\bf s_1},{\bf t_1}),\ldots,p_n({\bf s_n},{\bf t_n})$ is
{\em well-moded} iff the clause $p \la$
$p_1({\bf s_1},$ ${\bf t_1}),$ $\ldots,$ $p_n({\bf s_n},{\bf t_n})$ is
well-moded, where $p$ is any 
zero-ary predicate symbol.
\end{definition}

An example of a well-moded program and query is given in the next
subsection (Example \ref{exaconstr}).  In \cite{apt:emarchiori} the
persistence of the notion of well-modedness was proven; i.e.\/ {\em an
LD-resolvent of a well-moded query and a well-moded clause that is
variable-disjoint with it, is well-moded}.  Note that in a well-moded
query, the terms that occur in the input positions of the leftmost
atom are all ground.  Hence, as a consequence of the persistence of
well-modedness, we have that well-moded programs are data driven;
i.e.\/ {\em all atoms selected in an LD-derivation of a well-moded
query in a well-moded program contain ground terms in their input
positions}.

Next, we introduce the notion of simply-modedness \cite{apt:etalle}.
A family (or multiset)
of terms is called {\em linear} iff every variable occurs at
most once in it.

\begin{definition}[simply modedness]
~\\
A clause $p_0({\bf s_0},{\bf t_{n+1}}) \la$ 
$p_1({\bf s_1},{\bf t_1}),\ldots,p_n({\bf s_n},{\bf t_n})$ is
called {\em simply moded} iff 
${\bf t_1},\ldots,{\bf t_n}$ is a linear family of variables and
for $i \in [1,n]$:
\[
Var({\bf t_i}) \cap ( \bigcup_{j=0}^{i} Var({\bf s_j})) = \emptyset .
\]
A program is called {\em simply moded} iff every clause of it is simply
moded.
\\
A query 
$\la p_1({\bf s_1},{\bf t_1}),\ldots,p_n({\bf s_n},{\bf t_n})$ is
{\em simply moded} iff the clause $p \la$
$p_1({\bf s_1},{\bf t_1}),$ $\ldots,$ $p_n({\bf s_n},{\bf t_n})$ is
simply moded, where $p$ is any 
zero-ary predicate symbol.
\end{definition}

The program and query of Example~\ref{exaconstr} of the next subsection,
are simply moded.
The notion of simply modedness is persistent \cite{apt:etalle};
i.e.\ {\em an LD-resolvent of a simply moded query and a 
simply moded clause that is 
variable-disjoint with it, is simply moded}.
An atom is called {\em input/output disjoint} if the family of terms
occurring in its input positions has no variable in common with the 
family of terms occurring in its output positions.
As a consequence of the persistence of simply modedness, we have that
{\em all atoms selected in an LD-derivation of a simply moded query in 
a simply moded program 
are input/output disjoint and such that each of the output positions 
is filled in by a distinct variable}.

In \cite{apt:etalle}, it is argued that most programs are simply moded,
and that often non-simply moded programs can be naturally transformed into
simply moded ones.
In \cite{apt:etalle}, the class of simply moded, well-moded programs
and queries is shown to be unification-free, that is, in the execution,
unification can be replaced by iterated matching.

We want to note that our results for the class of simply moded
well-moded programs and queries carry over to the bigger class of
simply moded well-typed programs and queries such that the heads of
the program clauses are input safe.  We refer to 
\cite{VerbaetenCW99,Verbaeten_flops99},
where 
it is shown
why the results also hold for these programs and queries.  We also
refer to \cite{apt:etalle}, where this class of programs and queries
is introduced and shown to be unification-free.

\subsection{The Symbolic Condition for Quasi-Termination} 
\label{ssectconstraints}

We will now show how the rigid quasi-acceptability condition
(Proposition~\ref{propquswm})
is translated into a symbolic termination condition.
The symbolic condition is a system of constraints on the 
introduced symbols for norm,
level mapping and interargument relations (i.e.\ the functor, predicate
and extended predicate coefficients).
We first need the following concepts.

\begin{definition}
[measuring only/all input]
Let $C \subseteq B_P^E$. Let $\level{.}_s$ be a level mapping induced by
the symbol mapping $s$.
\\
We say that $\level{.}_s$ {\em measures only input positions in} $C$
iff
\begin{itemize}
\item
for every predicate $p/n$ occurring in $C$: if $i \in O_p$, then 
$s(p_i) = 0$,
\end{itemize} 
We say that $\level{.}_s$ {\em measures all input positions in} $C$
iff
\begin{itemize}
\item
for every predicate $p/n$ occurring in $C$: if
$i \in I_p$, then $s(p_i) \not= 0$, and

\item
for every functor $f/m$, $m>0$, 
occurring in an input position of an atom in $C$:
$s(f_i) \not= 0$ for all $i \in \{0,\ldots,m\}$.
\end{itemize} 
\end{definition}

The next lemma follows from the fact that, for a well-moded program $P$ 
and a set $S$ of well-moded queries, the input positions of atoms
in the call set $Call(P,S)$ are ground.

\begin{lemma} \label{proprigmod}
Let $P$ be a well-moded program and $S \subseteq B_P^E$ 
be a set of well-moded queries.
Let $\level{.}_s$ be a level mapping which measures only input positions
in $Call(P,S)$.
Then, $\level{.}_s$ is rigid on $Call(P,S)$.
\end{lemma}

The following lemma is a corollary of the fact that, for
a simply moded program $P$ and a set $S$ of simply moded queries,
the output positions of atoms in the call set $Call(P,S)$ 
consist of
distinct variables.

\begin{lemma} \label{propfinpartmod}
Let $P$ be a simply moded program and $S \subseteq B_P^E$ 
be a set of simply moded queries.
Let $\level{.}_s$ be a level mapping which measures all input positions
in $C \subseteq Call(P,S)$.
Then, $\level{.}_s$ is finitely partitioning on $C$.
\end{lemma}

In the following proposition from \cite{stef:constraints}, a condition
on a symbol mapping $s$ is given which ensures that the interargument
relations induced by $s$ are valid w.r.t.\ the norm induced by $s$.
We include
its proof
since it is essential for understanding the proposition.

\begin{proposition}
\label{validsymbolmapping}
Let $P$ be a program and $s$ a symbol mapping on ${\cal C}$.
If for each clause $H \leftarrow B_1,\cdots,B_n \in P$ it holds that
\[
s(\forall:[{\cal A}^s(B_1) \wedge \cdots \wedge {\cal A}^s(B_n) \Rightarrow
                {\cal A}^s(H) ])
\]
then for all $p/n$, $R_s^{p/n}$ is valid w.r.t.\ $\norm{.}_s$. 
\end{proposition}

\begin{proof}
The union of the relations 
\[
   R_s^{p/n} = \{(||t_1||_s),\cdots,||t_n||_s) \ |\ 
                 t_1,\cdots,t_n \in Term_P \mbox{ and }
                 s({\cal A}^s(p(t_1,\cdots,t_n))) \mbox{ holds } \}
\]
$p/n \in P$, define an interpretation of $P$ on the domain $\bbbn$.
The condition expresses that for this interpretation,
$T_P(I) \subseteq I$ holds.
Thus, the interpretation is a model and therefore
each $R^{p/n}_s$ is a valid interargument relation.
\qed
\end{proof}

Finally, we can formulate the rigid quasi-acceptability condition
in a constraint-based way.
We use Lemma \ref{proprigmod} and Lemma \ref{propfinpartmod},
and thus restrict
our results
to simply moded well-moded programs and queries.

\begin{proposition} \label{propquasi}
Let $P$ be a simply moded well-moded program.
Let $S \subseteq B_P^E$ be a set of simply moded well-moded queries.
Let $Tab_P \subseteq Pred_P$ be a tabling for $P$.
$P$ quasi-terminates w.r.t.\ $Tab_P$ and $S$
if
\\
there exists a symbol mapping $s$ such that
\begin{enumerate}
\item \label{itemonly}
$\level{.}_s$ measures only input positions in $Call(P,S)$:

\begin{itemize}
\item
for every predicate $p/n$ occurring in $Call(P,S)$:
 
if $i \in O_p$, then
$s(p_i) = 0$,
\end{itemize}

\item \label{itemall}
$\level{.}_s$ measures all input positions in $Call(P,S) \cap B^E_{Tab_P}$:

\begin{itemize}
\item
for every predicate $p/n$ occurring in $Call(P,S) \cap B^E_{Tab_P}$: 

if $i \in I_p$, then $s(p_i) \not= 0$, and

\item
for every functor $f/m$, $m>0$, 
occurring in an input position of an atom in $Call(P,S) \cap B^E_{Tab_P}$:

$s(f_i) \not= 0$ for all $i \in \{0,\ldots,m\}$,
\end{itemize} 

\item \label{itemvalidity}
all interargument relations induced by $s$ are valid
w.r.t.\ the norm $\norm{.}_s$
induced by $s$:

$\forall H \la B_1,\ldots,B_n \in P$:
\[
s(\forall:[{\cal A}^s(B_1) \wedge \cdots \wedge {\cal A}^s(B_n) \Rightarrow
{\cal A}^s(H) ]) ,
\]

\item \label{itemquasiacce}
the quasi-acceptability condition w.r.t.\ the norm, level mapping and
interargument relations induced by $s$ must hold:

$\forall H \la B_1,\ldots,B_n \in P$,
$\forall B_i, i \in \{1,\ldots,n\}$:
\[
s(\forall:[{\cal A}^s(B_1) \wedge \cdots \wedge {\cal A}^s(B_{i-1}) \Rightarrow
\level{H}^s \geq \level{B_i}^s ])
\]
and if
$Rel(H) \simeq Rel(B_i) \in NTab_P$
and $C_2(Rel(H),Rel(B_i))$ does not hold,
then
\[
s(\forall:[{\cal A}^s(B_1) \wedge \cdots \wedge {\cal A}^s(B_{i-1}) \Rightarrow
\level{H}^s > \level{B_i}^s ]).
\]
\end{enumerate}
\end{proposition}

\begin{proof}
This symbolic condition for quasi-termination is derived from the
rigid quasi-acceptability condition in a way analogous to the derivation
of the symbolic condition for LD-termination from the rigid
acceptability condition (see \cite{stef:constraints}).
In order for this article to be self-contained, we include the proof.
Suppose that there exists a symbol mapping $s$ satisfying the
above condition.
We prove that $P$ is rigid quasi-acceptable w.r.t.\ $Tab_P$ and $S$,
and hence that $P$ quasi-terminates w.r.t.\ $Tab_P$ and $S$.

We propose as a level mapping the level mapping $\level{.}_s$ induced 
by $s$ (based on the norm $\norm{.}_s$ induced by $s$).
Because $\level{.}_s$ measures only input positions in $Call(P,S)$,
we have by Lemma \ref{proprigmod} 
(and by the fact that $P$ and $S$ are well-moded)
that $\level{.}_s$ is rigid on $Call(P,S)$.
Also, because $\level{.}_s$ measures all input positions in 
$Call(P,S) \cap B^E_{Tab_P}$,
we have by Lemma \ref{propfinpartmod} 
(and by the fact that $P$ and $S$ are simply moded)
that $\level{.}_s$ is finitely partitioning on $Call(P,S) \cap B^E_{Tab_P}$.

Take any clause $H \la B_1,\ldots,B_n$ in $P$,
and any body atom $B_i$, $i \in \{1,\ldots,n\}$.
Let $\psi$ be a substitution such that 
$P \models B_1\psi, \ldots, B_{i-1}\psi$.
We prove that $\level{H\psi}_s \geq \level{B_i\psi}_s$
(the proof that $\level{H\psi}_s > \level{B_i\psi}_s$ in case
$Rel(H) \simeq Rel(B_i) \in NTab_P$ and $C_2(Rel(H),Rel(B_i))$ does not hold,
is analogous).
Condition \ref{itemquasiacce} of this proposition
holds for any instantiation of it, so
\[
s(\forall:[{\cal A}^s(B_1\psi) \wedge \cdots \wedge {\cal A}^s(B_{i-1}\psi) 
\Rightarrow
\level{H\psi}^s \geq \level{B_i\psi}^s ])  \qquad (*)
\]
holds.
Now, we prove that for any $1 \leq j \leq i-1$, $s(\forall: {\cal
A}^s(B_j\psi))$ holds.  By Proposition \ref{validsymbolmapping} and
condition \ref{itemvalidity} of this proposition, we have that, for
all $p/n$, $R^{p/n}_s$ is valid w.r.t.\ $\norm{.}_s$.
Then, since for all $1 \leq j \leq i-1$, $P \models B_j\psi$ holds, we
have that $s(\forall: {\cal A}^s(B_j\psi))$ holds.  So, by $(*)$ we
conclude that $s(\forall:\level{H\psi}^s \geq \level{B_i\psi}^s)$
holds, which implies that $\level{H\psi}_s \geq \level{B_i\psi}_s$
holds.
\qed
\end{proof}

Given a program $P$, a set of atoms $S$, and a tabling $Tab_P$, we can
set up a symbolic condition for quasi-termination using the above
proposition.
By solving the generated constraints, we get a demand-driven solution
for all the concepts involved in the termination analysis
(norm, level mapping and model).
More precisely, if a norm, level mapping and interargument relations of
the given generic forms exist such that the program can be proven to
quasi-terminate, then our generated set of constraints has these required
instances of the generic forms as a solution.

\begin{example} \label{exaconstr}
Let $P$ be the following program, computing the paths from a given
node to the reachable nodes in a given cyclic graph:
\[
\left\{
\begin{array}{lll}
edge(a,b) & \la &
\\
edge(b,a) & \la &
\\
path(X,Y,[Y]) & \la & edge(X,Y)
\\
path(X,Y,[Z|L]) & \la & edge(X,Z),\ path(Z,Y,L)
\end{array}
\right.
\]
Let $Tab_P = \{path/3\}$ and $S= \{path(a,Y,L)\}$.
Then, $P$ quasi-terminates w.r.t.\ $Tab_P$ and $S$.
We consider the following modes: $edge(In,Out), path(In,Out,Out)$.
Then the program $P$ and query $S$ are simply moded and well-moded.

We prove, using the constraint-based approach of
Proposition~\ref{propquasi}, that $P$ quasi-terminates w.r.t.\ $Tab_P$
and $S$.  We set up the constraints:

\begin{enumerate}
\item
First for the output positions of the predicates:
\[      s(edge_2) = 0, \quad s(path_2) = 0, \quad s(path_3) = 0.        \]

\item
For the input position of $path/3$:
\[      s(path_1) \not= 0.      \]

\item
We only need a linear size expression for 
the $edge/2$ predicate.  Its two clauses
both give rise to the following constraint:
\[
s(edge^e_1) 0  \geq s(edge^e_2) 0 + s(edge^e_0) .
\]
After simplification, we get: $s(edge^e_0) = 0$.

\item
The non-recursive clause for
$path/2$
gives rise to the following constraint:
\[
s(\forall X:[ path_1 X \geq edge_1 X ]) .
\]
This constraint has to hold for all possible values (in $\bbbn$) for $X$.
Hence, we derive the following constraint on the symbols
$path_1$ and $edge_1$:
\[      s(path_1) \geq s(edge_1).       \]
The recursive clause gives rise to the following two constraints:
\[
\begin{array}{l}
s(\forall X:[ path_1 X \geq edge_1 X ]) ,
\\
s(\forall X,Z:[ edge^e_1 X \geq edge^e_2 Z \Rightarrow
path_1 X \geq path_1 Z ]) .
\end{array}
\]
The first constraint is the same as the one for the non-recursive clause
and reduces to $s(path_1) \geq s(edge_1)$.
We refer to \cite{stef:constraints} for a general methodology 
for solving constraints as
generated by Proposition \ref{propquasi}.
Such constraints involve two types of variables: the symbolic 
coefficients for which we
aim to fix a symbol mapping and the universally quantified variables, which
express that the derived conditions should hold for any value of these;
the point is to eliminate the latter variables.
We briefly explain how the second constraint reduces to a system of constraints
on the symbolic coefficients only.
We first rewrite the second constraint into the following
equivalent form:
\[
s(\forall X,Z:[ edge^e_1 X - edge^e_2 Z \geq 0 \Rightarrow
path_1 X - path_1 Z \geq 0]) .
\]
Then the idea is to derive the right hand side as a positive linear
combination of the assumption in the left hand side.
We do this by subtracting the left hand side of the implication
from the right hand side, and by requiring that the resulting coefficients
of the variables are greater than or equal to $0$.
Doing so, we obtain the following constraints:
\[
s(path_1) - s(edge^e_1) \geq 0, \quad s(edge^e_2) - s(path_1) \geq 0 .
\]
\end{enumerate}

One solution to this system of constraints is ($\bullet$ stands for 
the list constructor):
\[
\begin{array}{l}
s(\bullet_0) = s(\bullet_1) = s(\bullet_2) = 0,
\\
s(edge_1)= 1, \quad s(edge_2) = 0,
\\
s(path_1) = 1, \quad s(path_2) = s(path_3) = 0,
\\
s(edge_1^e) = s(edge_2^e) = 1, \quad s(edge_0^e) = 0 .
\end{array}
\]

This gives us the following concrete norm and level mapping:
\[
\begin{array}{l}
\norm{t}_s = 0 \quad t \in U_P^E \quad (\norm{.}_s~\mbox{is the trivial norm}),
\\
\level{edge(t_1,t_2)}_s = \norm{t_1}_s ,
\\
\level{path(t_1,t_2,t_3)}_s = \norm{t_1}_s .
\end{array}
\]
The interargument relation for $\mathit{edge(t_1,t_2)}$ is 
$\norm{t_1}_s \geq \norm{t_2}_s$.

The rigid quasi-acceptability condition is satisfied using these concrete
norm, level mapping and valid interargument relation.
Hence, we have proven that $P$ quasi-terminates w.r.t.\ 
$\{\mathit{path}/2\}$
and $S$.
\end{example}

\section{Conclusions, Related 
Work and Topics for Future Research} 
\label{sectconc}

In this article we studied termination of 
tabled logic programs.
We introduced two notions of universal termination under a tabled
execution mechanism: quasi-termination and (the stronger notion of)
LG-termination.  We presented sufficient conditions (which are also
necessary in case the tabling is well-chosen) for quasi-termination
and LG-termination: namely quasi-acceptability and LG-acceptability.
We extended the applicability by presenting modular termination
conditions, i.e.\ conditions ensuring termination of the union $P \cup
R$ of two programs $P$ and $R$, where $P$ extends~$R$.
Finally, we investigated the problem of automatically proving
quasi-termination and LG-termination.  We showed that for simply
moded, well-moded programs, a sufficient condition for
quasi-termination and LG-termination can be given, which is formulated
fully at the clause level.  We pointed out how these sufficient
conditions can be automated by extending the constraint-based,
automatic approach towards LD-termination of~\cite{stef:constraints}.

Since all programs that terminate under LD-resolution, are
quasi-terminating and LG-termi\-na\-ting as well, verification of
termination under LD-resolution using an existing automated
termination analysis (such as those surveyed in e.g.~\cite{survey}) is
a sufficient proof of the program's quasi-termination and
LG-termi\-na\-tion.  However, since there are quasi-terminating and
LG-terminating programs, which are not LD-terminating, better proof
techniques can and should be found.  There are only relatively few
works studying termination under a tabled execution mechanism.
In \cite{stef:tabulation}, the special case where all predicates of
the program are tabled is considered and the two notions of universal
termination of a tabled logic program w.r.t.\ a set of queries is
introduced and characterised.  In \cite{pluemer:thesis}, in the
context of well-moded programs, Pl{\"u}mer presents a sufficient
condition for the bounded term-size property of programs, which
implies LG-termination.  Holst, in~\cite{Holst:FPCA91}, provides
another sufficient condition for quasi-termination in the context of
functional programming.

Our modular conditions for termination of tabled logic programs and
more precisely the modular conditions which incrementally construct a
level mapping for the whole program, are inspired by the modular
conditions for (S)LD-resolution as given by Apt and Pedreschi
in~\cite{apt:modular}.  More specifically, in \cite{apt:modular}, the
notions of semi-recurrent program (for SLD-resolution) and of
semi-acceptable program (for LD-resolution) are introduced, and
modular termination proofs are presented which are based on these
notions.

In the constraint-based approach towards quasi- and LG-termination, we
used mode information in the presentation of the sufficient
conditions.  In a recent article, \cite{EtalleBossiCocco98}, Etalle
\textit{et al} study how mode information can be used for
characterizing properties of LD-termination.  They define and study
the class of well-terminating programs, i.e., programs for which all
well-moded queries have finite LD-derivations.  They introduce the
notion of well-acceptability which is based on the concept of moded
level mapping; that is, a level mapping which only measures input
positions.  It is then shown that for well-moded programs,
well-acceptability implies well-termination.  Furthermore, it is
proven that for simply moded well-moded programs, the notions of
well-acceptability and well-termination are equivalent.

A topic for future research is to extend our results to \emph{normal}
logic programs executed under such a mixed tabled/non-tabled
execution.
Another topic, with an arguably more practical flavour, is to
investigate how the termination conditions presented here can form the
basis of a compiler that automatically decides on---or at least guides
a programmer in choosing---a tabling (i.e.\ a set of tabled
predicates) for an input program such that quasi-termination of the
program is ensured.  We plan to implement the constraint-based
technique for automatically proving quasi-termination and
LG-termination (note that a prototype implementation for automatically
proving LD-termination \cite{stef:constraints} exists).
Also, it remains to be studied how our results can be extended to
automatically prove quasi-termination and LG-termination for 
larger classes
of programs and queries (i.e.\/ for programs and queries which
are not simply moded,
well-moded).


\bibliographystyle{abbrv}
\bibliography{submTOCL}


\end{document}